\documentclass[12pt,letterpaper]{article}
\usepackage{jheppub}

\usepackage{epsfig,multirow,subfigure} 
\usepackage{amscd}
\usepackage[matrix,arrow,curve]{xy}
\usepackage{verbatim}
\usepackage{latexsym}
\usepackage{amsfonts,amsthm,amsmath,mathtools}

\newcommand{\Hyper}[1]{\Gamma_h\left( #1 \right)}
\newcommand{\be}{\begin{equation}}
\newcommand{\ee}{\end{equation}}
\newcommand{\bea}{\begin{eqnarray}}
\newcommand{\eea}{\end{eqnarray}}
 
\subheader{\begin{flushright}
UCSD-PTH-12-07\\ WIS/06/12-MAY-DPPA
 \end{flushright}}
\title{\centering
Refined Checks and Exact Dualities\\
 in Three Dimensions
}
\author[a]{Prarit Agarwal,}
\author[a]{Antonio Amariti,}
\author[b]{Massimo Siani}

\affiliation[a]{Department of Physics, University of California, \\
San Diego La Jolla, CA 92093-0354, USA}
\affiliation[b]{Department of Particle Physics and Astrophysics \\
Weizmann Institute of Science, Rehovot 76100, Israel}

\emailAdd{pagarwal@physics.ucsd.edu}
\emailAdd{amariti@physics.ucsd.edu}
\emailAdd{massimo@weizmann.ac.il}

\abstract{
We discuss and provide nontrivial evidence for a large class of dualities
in three-dimensional field theories with different gauge groups. 
We 
match the full partition functions of the dual phases
for any value of the couplings to underpin
our proposals.
We focus on two classes of models.
The first class, motivated by the AdS/CFT conjecture, 
consists of  necklace $U(N)$ quiver gauge theories with non chiral 
matter fields. We also consider orientifold projections and establish dualities
among necklace quivers with alternating orthogonal and symplectic groups.
The second class consists of  theories with tensor matter fields with 
free theory duals.  
In most of these cases the $R$-symmetry mixes with IR accidental symmetries
and we develop the prescription to include their contribution into 
the partition function and the extremization problem
accordingly.
}

\begin{document}

\maketitle
\section{Introduction}
Three dimensional dualities between supersymmetric field theories have been 
studied since a long time.
Some of them are similar to the four dimensional case of Seiberg duality,
like the Aharony duality \cite{Aharony:1997gp} and the Giveon-Kutasov 
one \cite{Giveon:2008zn}.

More recently, new nonperturbative techniques have been used to gain more
insights into aspects of three-dimensional field theories. In particular, the exact partition function
of any ${\cal N}\ge 2$ superconformal field theory reduces to a matrix model for any value of
the coupling constants \cite{Kapustin:2009kz,Jafferis:2010un,Hama:2010av}, 
and gives information about physical quantities of the given model
\cite{Suyama:2009pd,Herzog:2010hf,Martelli:2011qj,Cheon:2011vi,Jafferis:2011zi,Amariti:2011uw,
Minwalla:2011ma,Amariti:2011da,Amariti:2011jp,Gang:2011jj,Amariti:2011xp,
Drukker:2010nc,Amariti:2012tj}
that can be compared with previous results
\cite{Bagger:2006sk,Bagger:2007jr,Gustavsson:2007vu,Bagger:2007vi,Aharony:2008ug,
Gaiotto:2007qi,Avdeev:1992jt,
Jafferis:2008qz,
Martelli:2008si,
Hanany:2008cd,
Hanany:2008fj,
Ueda:2008hx,
Imamura:2008nn,
Franco:2008um,
Hanany:2008gx,
Bianchi:2009ja,
Amariti:2009rb,
Franco:2009sp,
Bianchi:2009rf,
Davey:2009sr}.

Moreover, one can also compare the partition functions of two field theories that are conjectured to
describe dual phases of the same superconformal fixed point, thus providing a nontrivial check of
the duality. Showing that both sides share the same partition function is   non trivial .
One can consider different limits.
Seiberg-like dualities for theories with at least ${\cal N}=3$ supersymmetry have been
considered in \cite{Kapustin:2010mh,Kapustin:2010xq}. With lower supersymmetry, the partition
function is considerably more complicated. In the large-$N$ limit, one can use the saddle point
approximation and successfully study infinite classes of theories which involve an arbitrary product
gauge group \cite{Amariti:2011uw,Gulotta:2011vp}.
For finite values of the gauge group rank and Chern-Simons  (CS) level one can exploit the following
observation.
Exact results can be also obtained by generalizing the three-dimensional space on which the
theory is defined to a squashed three-sphere, which enjoys a $U(1)^2$ subgroup of the isometry
group $SU(2)^2$ of $S^3$. The localized partition function on this space can be written in terms
of hyperbolic functions \cite{Hama:2011ea}. A review of their properties is given in \cite{fvdb}, and in
appendix \ref{vdb section}, and they have revealed themselves
very useful to give further evidence to a large
class of dualities \cite{Willett:2011gp,Kapustin:2011gh,Benini:2011mf,
Niarchos:2012ah}.

In most of these cases a single gauge group has been considered,
but in principle one
can use the same approach to match exact results for physical quantities among
dual phases of theories describing generic configurations of M$2$ branes.

In this paper we are interested in different classes of dualities. Some of these have been considered in the framework
of the large-$N$ approximation of the partition function in \cite{Amariti:2011uw,Gulotta:2011vp}. However,
this limit does not catch an important subtlety of the duality transformation.
 If one starts with a product
of unitary gauge groups $\prod U(N)_{k_i}$ in the electric theory and performs a duality transformation
on the group $i$ the resulting dual gauge group contains a factor $U(N+|k_i|)$. At the leading order in
a large-$N$ expansion, this dependence upon the CS level $k$ does not play any role.
We drop the large-$N$ limit and provide nontrivial evidence for this duality to hold at any value of
$N$ and the $k$'s in section \ref{sec:dualities}. We also consider other models, which can be derived
as the low energy theories living on the worldvolume of intersecting D-branes and orientifold O-planes, their
dual phases and match the finite-$N$ partition function for them.

Another interesting set of dualities recently proposed in \cite{Jafferis:2011ns} and extended
in \cite{Kapustin:2011vz}
can be studied by computing the partition function on the squashed three-sphere. In these cases we re-derive 
some of the known results by applying the exact calculations of \cite{fvdb} and we compare with  known dualities.

In these cases one has to pay attention to infrared accidental symmetries. Indeed in some cases
the exact computation shows that some theories look dual to   free theories in which the  scaling
dimensions of the gauge invariant operators
are not consistent with the free theory value.
  A proper modification of the extremization principle, to account for the mixing of accidental symmetries with the $R$-symmetry,
is necessary for the calculation of the exact $R$-charge.

The paper is organized as follows. In section \ref{sec:partfun} we review the rules to write the all-loop
partition function on a squashed three-sphere, and show how it can be written in terms of hyperbolic
functions. We also list a few basic properties of the hyperbolic functions. In section \ref{sec:models}
we review some of the classes of models we are interested in. We describe how they can be embedded
in a type IIB setup, and how the duality transformations follow from this embedding. We consider
theories with unitary, orthogonal and symplectic factors in the product gauge group. The dualities
are proved for any value of the ranks and CS levels in section \ref{sec:dualities} through
the matching of the partition functions on both sides. Models with free field theory duals will be considered
in section \ref{sec:free}, where we also raise the problem of accidental symmetries which we further
describe in section \ref{sec:acc}. Open problems and hints for future work are discussed in
section \ref{sec:open}. We include some appendices which contain technical details.

\section{The partition function on a squashed three sphere} \label{sec:partfun}

Localization
has allowed to reduce the partition function 
of any $\mathcal{N}=2$ three dimensional supersymmetric theory 
on a three sphere $S^3$ \cite{Jafferis:2010un,Hama:2010av}.
A further refinement \cite{Hama:2011ea} involves two different squashed spheres $S_b^{3}$:
One of them preserves an $SU(2)\times U(1)$ 
isometry, but in this case the localization does not give any new result,
the other one, which will be very useful in this paper, preserves an $U(1)^2$ 
isometry.
The partition function on the latter squashed sphere $S_b^{3}$ 
for a CS matter theory with gauge group $G$ is 
\begin{eqnarray} \label{eq:partfunct}
\mathcal{Z}_{S_b^3}&=&    \int_{\mathbb T^{\text{rk} (G)}}
\prod_{i=1}^{\text{rk}(G)} d x_i e^{ i \pi k  Tr_F x^2}
\det_{~~~~adj} \left( 
\sinh\left( \pi b \rho_\alpha(x)\right)
\sinh\left( \pi b^{-1} \rho_\alpha(x)\right)
\right) \nonumber \\
&\times&\prod_{\rho \in r}S_b\left(\frac{i}{2} (b+b^{-1}) (1-\Delta_{r}) - \rho_r(x)\right)
 \end{eqnarray}
  where $\Delta_{r}$ is the scaling dimension (which in three dimensions coincides with the $R$-charge) of a
chiral matter
field in the representation $r$, $\rho_r$ are the weights of the representation $r$,  and $\rho_ \alpha$ 
are the roots of the gauge groups $G$.
The various factors in the integrand in  (\ref{eq:partfunct}) 
correspond to the contribution from the CS term, the vector multiplet and the matter superfields 
(in the representation $r$) respectively.
The function 
$S_b$ is the double sine function defined as
\begin{equation}
S_b\left(\frac{i}{2} (b+b^{-1}) (1-\Delta_{r}) - \rho(x)\right)
=
\prod_{n_1,n_2 \geq 0}^{\infty}
\frac{n_1 b + n_2 b^{-1} + \frac{b+b^{-1}}{2} + i \rho(x) +\frac{b+b^{-1}}{2}(1-\Delta_r) }
{n_1 b + n_2 b^{-1} + \frac{b+b^{-1}}{2} - i \rho(x) -\frac{b+b^{-1}}{2}(1-\Delta_r) }
\end{equation}
The limit $b=1$ corresponds to the round sphere considered in \cite{Jafferis:2010un,Hama:2010av}.
In that case the double sine reduces to
\begin{equation}
S_1\left(i(1-\Delta_{r}) - \rho(x)\right) \equiv S_1(i z)= e^{l(z)}
\end{equation}
where $l(z)$ is defined such that its derivative is $-\pi z \cot(\pi z)$.

The partition function on the squashed sphere is more complicated than the corresponding one
on the round sphere. However, since the double sine function can be identified with the
hyperbolic Gamma function \cite{Ruijsenaars}, we can exploit the recent work by mathematicians
which provide us with exact results for the integral involved in physical computations \cite{fvdb}.
In the following we introduce the basic definitions relevant for this paper,  and provide
more technical details
to appendix \ref{vdb section}.

 \subsection{Hyperbolic functions}
 
 We start by introducing the \emph{periods} $\omega_1$ and $\omega_2$, that in this case are identified with
 \begin{equation}
 \omega_1 = i b \quad,\quad
 \omega_2 = i b^{-1} \quad,\quad
 \omega = \frac{\omega_1+\omega_2}{2}
 \end{equation}
 The double sine function in terms of $\omega_1$, $\omega_2$ and $z$ becomes
\begin{equation} 
S(-i z;-i\omega_1,-i\omega_2)
=
\prod_{n1,n2 \geq 0}^{\infty}
\frac{(n_1+1)\omega_1+(n_2+1)\omega_2- z}
{n_1 \omega_1+n_2 \omega_2+z}
\end{equation}
This corresponds to the hyperbolic gamma function 
$\Gamma_h (z;\omega_1,\omega_2)\equiv\Gamma_h(z)$ 
first defined in  \cite{Ruijsenaars}.
  This function satisfies the difference equations
 \begin{equation} \label{diffeq}
 \Gamma_h(z+\omega_1) = 2\sin\left(\frac{\pi z}{\omega_2}\right) \Gamma_h(z)
 \quad
 ,
 \quad
  \Gamma_h(z+\omega_2) = 2\sin\left(\frac{\pi z}{\omega_1}\right) \Gamma_h(z)
  \end{equation}
and the reflection formula
\begin{equation}\label{refle}
\Gamma_h(z+\psi_1)\Gamma_h(\psi_2-z)=1\quad \quad  \text{if}\quad  \quad \psi_1+\psi_2=2 \omega 
\end{equation}
Other useful identities are
\begin{equation}\label{otherid}
\Gamma_{h}(\omega)=1\,,\, \Gamma_{h}\left(\frac{\omega_1}{2}\right)= \Gamma_{h}\left(\frac{\omega_2}{2}\right)=\frac{1}{\sqrt{2}}
\,,\, 
\Gamma_h\left(\omega + \frac{\omega_1}{2}\right)= \Gamma_{h}\left(\omega+\frac{\omega_2}{2}\right)=\sqrt{2}
\end{equation}
and
\begin{eqnarray}
\label{usefullater}
\Gamma_h(2 z) = \Gamma_h\left(z\right) \Gamma_h\left(z+\frac{\omega_1}{2}\right) \Gamma_h\left(z+\frac{\omega_2}{2}\right)
\Gamma_h(z+\omega) \\
\end{eqnarray}
By combining (\ref{diffeq}) and (\ref{refle})
one has
\begin{equation}
\Gamma_h(\pm z)\equiv
\Gamma_h(z) \Gamma_h(-z) 
=
\frac{\Gamma_h(z+\omega_1)\Gamma_h(\omega_2-z)}{4
\sin\left( \frac{\pi z}{\omega_1}\right)
\sin\left(- \frac{\pi z}{\omega_2}\right)}
=-
\frac{1}{4
\sin\left( \frac{\pi z}{\omega_1}\right)
\sin\left( \frac{\pi z}{\omega_2}\right)}
\end{equation}
which correspfonds to the one loop contribution of the vector multiplet in (\ref{eq:partfunct}).
The final expression for the partition function in terms of the 
hyperbolic gamma function is
\begin{eqnarray}
\mathcal{Z}(\Delta_R, \omega_1,\omega_2)=
\frac{1}{\sqrt{(-\omega_1\omega_2)^n} {\cal W}}
  \int
\prod_{i=1}^{n} d u_i e^{\frac{- i \pi k}{\omega_1 \omega_2}  x_i^2}
\frac{\prod_{\rho_r \in R}\Gamma_h( \rho_r(x)+ \omega \Delta_R)}
{\prod_{\rho_{\alpha} \in \alpha^{(+)}} \Gamma_h(\pm \rho_\alpha(x_i))}
 \end{eqnarray}
where $\cal W$ is the dimension of the Weyl subgroup and $n$ is the rank of the
gauge group.
Many exact results concerning these integrals have been studied in \cite{fvdb}.
To deal with the notations there we define the 
functions $c(x)$ and $\zeta$ as
\begin{equation}
c(x)\equiv \exp\left(\frac{i \pi x}{2 \omega_1 \omega_2}\right)
\quad,\quad
\zeta=e^{\frac{i \pi(\omega_1^2+\omega_2^2)}{24 \omega_1\omega_2} }
\end{equation}
in terms of which the CS contribution at level $k=- \frac{t}{2}$ is
\begin{equation}
\exp\left({\frac{ i \pi t}{2\omega_1 \omega_2}  x_i^2}\right) = c(t x_i^2)
\end{equation}
Also notice that 
in the $S^3$ limit,
$\omega_1=\omega_2=i$, we obtain 
 $\log\left(\Hyper{z}\right)=l(1+i z)$
 which is the one loop  contribution of matter fields  computed in
 \cite{Jafferis:2010un}.

\section{Families of quiver gauge theories and M$2$ branes} \label{sec:models}

In this section we survey the  classes of models 
dual to M$2$ branes on Calabi-Yau fourfold that we will
be interested in.
These models have been deeply investigated in \cite{Jafferis:2008qz,
Martelli:2008si,
Hanany:2008cd,
Hanany:2008fj,
Ueda:2008hx,
Imamura:2008nn,
Franco:2008um,
Hanany:2008gx,
Amariti:2009rb,
Franco:2009sp,
Davey:2009sr}.

Each one can be understood in the framework of type IIB 
SUGRA compactified on a circle.
The low energy brane dynamics is described by
the worldvolume theory living in the $2+1$ infinite directions of some D$3$ brane
suspended between pairs of  $(1,p_i)$ fivebranes.
The latter picture also provides us with a representation in terms of
quiver diagrams, according to which we associate a node to each gauge group and an arrow
to each matter field.
We distinguish two types of arrows: one which connects two 
distinct nodes is associated to bifundamental matter fields, while one that
has both its endpoints on the same node represents a chiral field in the adjoint representation.

In the three-dimensional case, in addition to the above information we also have to provide the CS 
levels. In the type IIB picture, they are given by the difference $(p_i-p_{i-1})$.
 From a purely field theoretical point of view, our only constraint will be that they add up
to zero.

Finally, we will let the gauge group factors to be either the unitary, orthogonal or symplectic group
(i.e. we also consider  cases with O$3$ planes in the brane construction).

\subsection{Unitary groups}

We take type IIB string theory compactified on a circle, which we parametrize with the $x_6$ coordinate.
The worldvolume theory of a stack of $N$ D$3$ branes wrapped on the circle is described by a $U(N)$ gauge
theory in three dimensions. If the D$3$'s intersect $g$ NS$5$ extended along the D$3$ worldvolume but
not around the circle, the gauge group contains $g$ $U(N)$  factors. The introduction of the CS
terms is achieved by replacing the NS$5$ with a tilted bound state of NS$5$ and $p_i$ D$5$, dubbed $(1,p_i)$
fivebrane. We refer to table \ref{tab:branes} for the precise definition of the embedding. The $(0,1,2)$
directions represent the three-dimensional spacetime, with $x_6$ compact.
 \begin{table}[ht]
\centering
 \begin{tabular}{c||cccccccccc}
 brane & 0&1&2&3& 4&5&6&7&8&9\\
 \hline
 D$3$ & X&X&X&& &&X&&&\\
NS$5_{\alpha}$ & X&X&X&X& X&X&&&&\\
NS$5_{\beta}$  & X&X&X&X& &&&&X&X\\
   D$5_\alpha$ & X&X&X&&X &X&&X&&\\
  D$5_\beta$ & X&X&X&& &&&X&X&X\\
   \end{tabular}
\caption{Type IIB embedding of low energy CS field theories.}
\label{tab:branes}
  \end{table}
The $\alpha$-th NS$5$ brane, $\alpha=1,\ldots,a$, combines with 
the $Q_\alpha$ D$5_{\alpha}$ branes 
to give a $(1,Q_{\alpha})$-fivebrane stretched along the 
$012[37]_{\theta_{\alpha}}45$  direction.
The $\beta$-th NS$5$ brane, $\beta=1,\ldots,b$, combines with 
the $P_\beta$ D$5_{\beta}$ branes 
to give a $(1,P_{\beta})$-fivebrane stretched along the 
$012[37]_{\theta_{\beta}}89$  direction. For specific
 values of the
angles $\theta_\alpha$ and $\theta_\beta$ determined by $Q_\alpha$ and $P_\beta$,
the supersymmetry is enhanced to
${\cal N} \ge 3$. We will consider generic configurations, so our results will be also valid
when this enhancement does occur.
The fivebranes are chosen to be placed in the following order:
first we put $b-a$ $(1,P_\beta)$ fivebranes 
on the circle and then we alternate the remaining 
$a$ $(1,P_\beta)$ and the $a$ $(1,Q_\alpha)$.

The $N_i$ D$3$-branes stretched between each pair of $(1,p_i)$ give rise to a $U(N_i)_{k_i}$ gauge group
in the quiver. Each $(1,p_i)$ is associated to a pair of bifundamental chiral fields in the $(N_i,\bar N_{i+1})$
representation.
In addition, we have an adjoint chiral field for each consecutive pair
of $(1,p_i)$ of the same type.
The resulting field theory is a $ \prod_{i=1}^{g} U(N_i)_{k_i}$
gauge theory, where $k_i$ represents the CS level of the
$i$-th group and $g=b+a$.
The levels are given by the relation 
$k_i=p_i-p_{i+1}$ which also implies $\sum k_i=0$.
In the quiver representation we have the first $b$ nodes with adjoint matter and the last $a$ without adjoints;
every pair of consecutive nodes is connected by a pair  of bifundamental and anti-bifundamental fields.
We also obtain the following superpotential
\bea
W=X_{1,1}X_{1,a+b}X_{a+b,1} + \sum_{i=1}^{b-a}X_{i,i} X_{i,i+1} X_{i+1,i}+ \sum_{i=b-a+1}^{a+b} X_{i,i-1} X_{i-1,i} X_{i,i+1} X_{i+1,i} 
\eea
where $X_{i,j}$ indicates a bifundamental field connecting nodes $i$ and $j$ and $X_{i,i}$ corresponds to an
adjoint of the node $i$.
Globally, the brane construction and thus the field theory preserves $\mathcal{N}=2$ supersymmetry.

\subsubsection{Duality}

The above brane picture allows us to describe Seiberg-like dualities in an unified way, through the
Hanany-Witten transition \cite{Hanany:1996ie}.
Consider two consecutive, non-parallel, $(1,p_{i})$ fivebranes and move one towards the other
until they cross and exchange their positions along the $x_6$ direction.
Quantum charge conservation requires the creation of $|P_{\beta_{b+i}}-Q_{\alpha_i}|=k_{b+i}\,\,$ 
D$3$ branes on top of the existing $N_{b+i}$ ones.

Correspondingly, in the low energy field theory the $i$-th gauge factor changes its rank
from $N_i$ to $N_i+|k_i|$, and because the fivebrane charges and order determine the CS levels,
the latter also undergo the following shift
\bea
k_{i-1}&=&p_{i-1}-p_i \,\,\to\,\, k_{i-1}^\prime = p_{i-1}-p_{i+1}=k_{i-1}+k_{i} \nonumber \\
k_{i}&=&p_{i}-p_{i+1} \,\,\to\,\, k_{i}^\prime = p_{i+1}-p_{i}=-k_{i}  \\
k_{i+1}&=&p_{i+1}-p_{i+2} \,\,\to\,\, k_{i+1}^\prime = p_{i}-p_{i+2}=k_{i+1}+k_{i} \nonumber
\eea
Note that the sum of all the CS levels is preserved in this process. The local nature of the Hanany-Witten
transition is reflected in the field theory by the fact that only one gauge group and its first neighbors
go through a change.
Finally the superpotential locally changes as\footnote{Actually
 also the nodes b-2 and b-2 are involved, because there are
two extra terms
$X_{b-2,b-1} Y_{b-1,b-1} X_{b-1,b-2}$ and
$X_{b+2,b+1} Y_{b+1,b+1} X_{b+1,b+2}$ in the superpotential. 
We can skip this contribution in our analysis
because  the $R$-charges of
$X$ fields are not affected.}
\begin{eqnarray}
\widetilde W &=& Y_{b+i-1,b+i-1}Y_{b+i-1,b+i}Y_{b+i,b+i-1}+
Y_{b+i,b+i-1}Y_{b+i-1,b+i}
Y_{b+i,b+i+1}Y_{b+i+1,b+i}\nonumber \\
&+&
Y_{b+i+1,b+i+1}Y_{b+i+1,b+i}Y_{b+i,b+i+1}
\end{eqnarray}
Notice that the dual theory also contains two new adjoint fields.
Thanks to the above superpotential, the two models have the same moduli space and are conjectured to
be dual to each other in the deep infrared. Also notice that nowhere did we use the fact that in this example
the electric ranks of the
gauge groups are equal to each other. Thus the same argument can be straightforwardly applied
to a product of arbitrary unitary groups.

The duality above extends the Kutasov-Giveon duality \cite{Giveon:2008zn} for 
three dimensional supersymmetric gauge theories with CS terms.
Nontrivial checks are required in order to validate the whole picture provided above. In fact,
there exist two limits where such checks have been given. One is the large $N$ limit \cite{Amariti:2011uw}.
In this case, the dual
gauge group can be safely taken to be the original one, because any difference in the ranks due to the CS levels
is subleading. Notice that, in general, this is a nontrivial statement.\footnote{We are grateful
 to Claudius Klare and Alberto Zaffaroni for discussions on this point.}

The second limit corresponds to the case of finite $N$ with $\mathcal{N}\geq 3$.
Only partial results have been studied in this limit. For instance, when 
$a=b=1$ the model is the ABJM model which enjoys
$\mathcal{N}=6$ supersymmetry. In that case the analysis becomes much simpler and many checks
have been provided.
In fact, beside the moduli space matching, there is no check for models with $g>2$ gauge group factors
and ${\cal N}=2$ supersymmetry. The main difficulties in this case are due to the nontrivial anomalous dimensions
of the fields. We will see how we can identify the scaling dimensions of the fields on the two sides
of the duality so that the two partition functions agree even for $g>2$ and for arbitrary ranks.
We will also argue that the map we will
describe preserves extremization of the partition function with respect to scaling dimensions themselves.

\subsection{Orthogonal and symplectic groups: the orientifold}
\label{subsect:ori}
While we focused on unitary gauge groups in the above subsection, more general models can be derived
from the same type IIB picture above. An immediate extension includes adding orientifold O$3$ planes on
top of the D$3$ branes, which we employ in the following. This construction does not break any residual
supersymmetry, so we will end up with ${\cal N}\ge 2$ theories
\cite{Hosomichi:2008jb,Aharony:2008gk}.\footnote{Other
 orientifold constructions that break supersymmetry have been investigated in 
\cite{Armoni:2008kr,Forcella:2009jj}.}

For simplicity, we restrict to the class of theories with $a=b$.
Under the orientifold projection, the $(1,p_i)$ fivebranes which intersect the O$3$ are
identified with their own image while the projection does not act on the D$3$ branes.
There are four kind of O$3$ planes, named O$3^{\pm}$ and $\widetilde {O 3}^{\pm}$, that differ,
among other, by the amount of D$3$ brane charge they carry, and by the resulting worldvolume theory
gauge group they lead to.	We summarize the different cases in table \ref{tab:O3}.
\begin{table}
 \begin{center}
\begin{tabular}{c||c|c}
Type & Charge &Group\\
\hline
O$3^{+}$& $\,-\frac{1}{4}$ & $SP(2N)$\\
O$3^{-}$&$~~~\frac{1}{4}$& $SO(2N)$\\
$\widetilde{\text{O}3}^{+}$&$~~~\frac{1}{4}$& $SP(2N)$\\
$\widetilde{\text{O}3}^{-}$&$~~~\frac{1}{4}$& $SO(2N+1)$
\end{tabular}
\end{center}
\caption{O$3$ planes, their D$3$ brane charge and the corresponding gauge group.}
\label{tab:O3}
\end{table}
A $(1,p_i)$ fivebrane which intersects the orientifold plane switches its type according
to the following rule: 
If $p_i$ is even we have $(\text{O}3^{+},\widetilde{\text{O}3}^{+})\leftrightarrow
 (\text{O}3^{-},\widetilde{\text{O}3}^{-})$
otherwise if $p_i$ is odd we have $(\text{O}3^{\pm} \leftrightarrow \widetilde{\text{O}3}^{\mp})$.
We restrict to the case of $p_i$ even and make this explicit by considering $(1,2 p_i)$ instead.
According to the general discussion on the brane engineering of CS
matter theories above, all the CS terms will be even too.

\begin{figure}[htbp]
\begin{center}
\includegraphics[width=15cm]{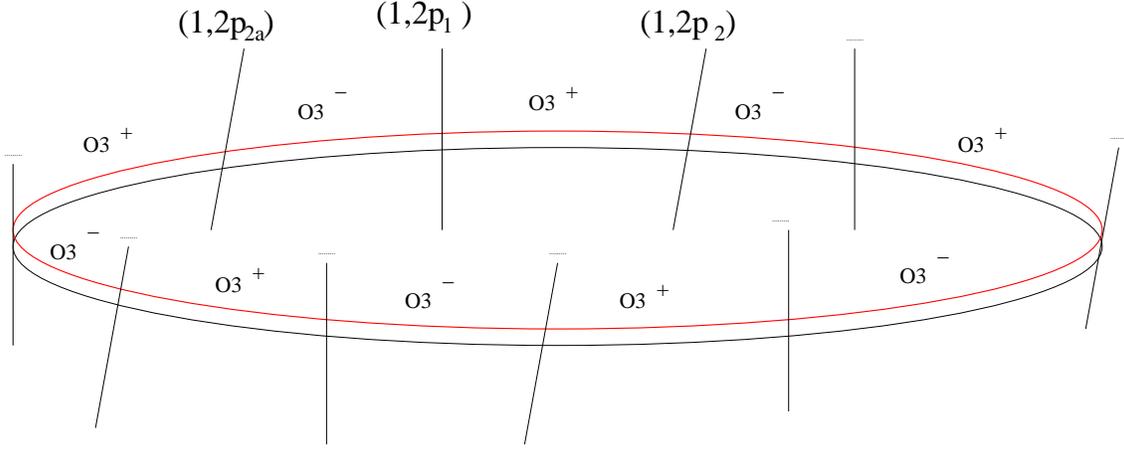}
\end{center}
\caption{A type IIB embedding of orthogonal and symplectic field theories via O$3^{\pm}$ planes on a stack of D$3$.
}
\label{FigOri}
\end{figure}

It is then clear that the gauge group will include alternating factors of orthogonal and symplectic groups. Their
ranks are given by the choice of the O$3$ planes, namely we obtain a chain of
$SO(2N)_{2 k_i} \times SP(2N)_{k_j}$ factors for alternating O$3^{+}$ and O$3^-$ planes, and of
 $SO(2N+1)_{2 k_i} \times SP(2N)_{k_j}$ factors in the $\widetilde{\text{O}3}^{\pm}$ case
(with the convention $SP(2)\simeq SU(2)$).\footnote{Observe that the level of the SP group $k$ is integer.
For this reason taking an odd number of  D$5$  in the fivebrane
is quantum mechanically inconsistent, because we would get a semi-integer CS level.
  Moreover in the CS contribution 
 to the partition  function there will be an extra factor of $2$ for the SP cases,
due to normalization of the generators \cite{Willett:2011gp}.}
An example of such construction is given 
in figure \ref{FigOri} for the case with O$3^{\pm}$.
 The fields are projected such that every pair of bifundamental and anti-bifundamental 
 becomes a single field in the fundamental of both the $SP$ and $SO$ node.
 By starting with $X_{i-1,i}$ and $X_{i,i-1}$ one ends up with a single field $X_{i-1,i}$
 when we fix the left to right convention on the indices.
The superpotential is
\begin{equation}
W = \left(X_{i-1,i} \cdot 
X_{i,i+1}\right)^2
\label{eq:SOW}
\end{equation}
 where the products  are appropriately  taken in the $SP$ and/or $SO$ case.

 \subsubsection{Duality}

We again apply the brane creation effect when two fivebranes cross each other to derive the rules for the
low energy field theory duality. The steps are in close analogy with the ones above, with the charge of
the O$3$ plane properly taken into account.

Because the duality only acts locally on the quiver, we can isolate the node over which we perform the
fivebrane exchange and collect the changes in the gauge group and CS level of itself and its neighbors
as follows: suppose we apply the duality on the node $A$ which locally looks like
\begin{center}
\begin{tabular}{c||c||c}
A-1&  A& A+1\\
\hline
$SO(2N)_{2k_{A-1}}$&$SP(2N)_{k_A}$&$SO(2N)_{2k_{A+1}}$\\
$SO(2N+1)_{2k_{A-1}}$&$SP(2N)_{k_A}$&$SO(2N+1)_{2k_{A+1}}$\\
$SP(2N)_{k_{A-1}}$&$SO(2N)_{2k_A}$&$SP(2N)_{k_{A+1}}$\\
$SP(2N)_{k_{A-1}}$&$SO(2N+1)_{2k_A}$&$SP(2N)_{k_{A+1}}$
\end{tabular}
\end{center}
with superpotential \eqref{eq:SOW}. Then the dual theory is locally given by
\begin{center}
\begin{tabular}{c||c||c}
A-1&  A& A+1\\
\hline
$SO(2N)_{2k_{A-1}+2k_A}$&$SP(2(N+|k_A|-1)_{-k_A}$&$SO(2N)_{2k_{A+1}+2k_A}$\\
$SO(2N+1)_{2k_{A-1}+2k_A}$&$SP(2(N+|k_A|-1))_{-k_A}$&$SO(2N+1)_{2k_{A+1}+2k_A}$\\
$SP(2N)_{k_{A-1}+k_A}$&$SO(2(N+k_A-1))_{-2k_A}$&$SP(2N)_{k_{A+1}+k_A}$\\
$SP(2N)_{k_{A-1}+k_A}$&$SO(2(N+|k_A|)+1)_{-k_A}$&$SP(2N)_{k_{A+1}+k_A}$
\end{tabular}
\end{center}
with all the remaining nodes in the quiver unchanged and dual superpotential given by
\begin{equation}
\tilde W = Y_{A-1,A-1} \cdot Y_{A-1,A}^2+ Y_{A-1,A}^2 Y_{A,A+1}^2 +Y_{A+1,A+1}\cdot  Y_{A+1,A}^2 
\end{equation}
These dualities fit with the ones proposed in \cite{Kapustin:2011gh} for the case without the quiver structure, 
and with the ones for the case of two gauge groups and higher supersymmetry \cite{Aharony:2008gk}.

In the following we will show that the partition function is preserved at finite $N$ for all of these dualities.

\section{Exact results for the dualities} \label{sec:dualities}

In this section we  evaluate the exact  partition function on a squashed three sphere
of the above models and
provide further evidence for the dualities. We review the identities we use in Appendix \ref{vdb section}
and also refer to \cite{fvdb} for more details.
Because the duality
only acts on the local structure of the quiver, we
can restrict ourselves to the the subset of variables which undergo
the duality transformation. In other words, we explicitly write only the integration variables corresponding
to the gauge group factor we are performing the duality on.

\subsection{Duality in $U(N)_k$ non-chiral  quivers}

\label{dualityUN}

In this case the large $N$ partition function have been studied in
\cite{Drukker:2010nc,Herzog:2010hf,Martelli:2011qj,Jafferis:2011zi,Cheon:2011vi,
Suyama:2009pd,Amariti:2011uw},
and the agreement between dual phases have been checked in this limit in
\cite{Kapustin:2010mh,Kapustin:2010xq,Gulotta:2011vp,
Amariti:2011uw,
Amariti:2012tj}.
Here we provide the agreement at finite $N$.

\begin{figure}
\begin{center}
\includegraphics[width=10cm]{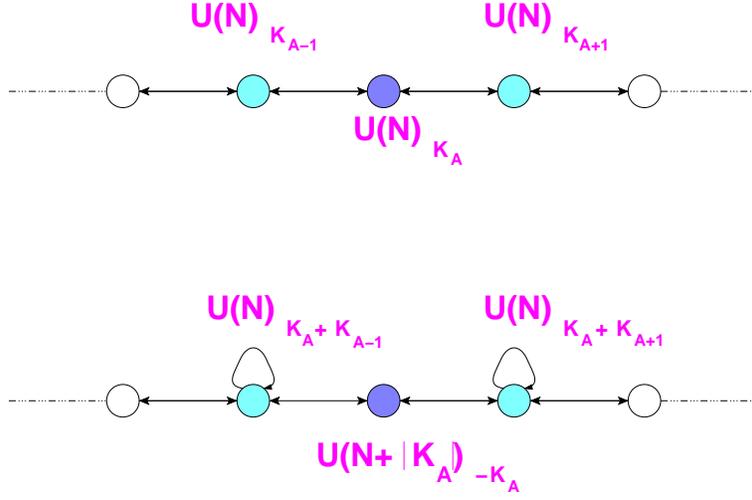}
\end{center}
\caption{Dual phases describing a stack of M$2$ branes probing a Calabi-Yau fourfold.}
\label{Figu}
\end{figure}

In terms of the hyperbolic functions defined in section \ref{sec:partfun}, the partition function for models with
only unitary gauge groups can be written in a very compact way.
The matter content and local quiver structure are represented in figure \ref{Figu}, where we
used the letter $A$ to label the gauge group factor over which we perform the duality. From the top figure
we read the 
relevant contribution to the partition function involved in the duality
as
%
%
%
%
%
\small
\bea \label{elpt}
\mathcal{Z}_e &=& \frac{1}{\sqrt{\left(-\omega_1\omega_2\right)^N} N!}
\displaystyle\int \! \frac{\displaystyle \prod_{J=A-1}^{A}  \prod_{i,j=1}^{N}  \prod_{\eta=\pm 1} \Gamma_h\left(\eta \left( x_J^{(i)} -x_{J+1}^{(j)} \right) + \omega \Delta_{J,J+1}^{(\eta)} \right) }
{\displaystyle \prod_{i<j}^{N} \Gamma_h \left(\pm \left( x_{A}^{(i)}-x_{A}^{(j)}\right)\right)}
\nonumber \\
&\times&\prod_{J=A-1}^{A+1}  \prod_{i=1}^{N} c\left(-2 k_J {x_J^{(i)}}^2\right) \prod_{i=1}^{N}d x_{A}^{(i)} 
\eea
\normalsize
where the round sphere corresponds to the limit $\omega_1=\omega_2=\omega=i$. Our aim is to write \eqref{elpt}
in a form that can be interpreted as the partition function of the dual theory described in section \ref{sec:models}.
We find it is useful to define the following shorthand notation
\bea
\begin{array}[t]{cc}
\begin{array}{l}
\Delta_{J,J+1} = \Delta_{J,J+1}^{(+)} \\
\mu_{A+1}^{(i)}=x_{A+1}^{(i)} + \omega \Delta_{A,A+1}^{(-)} \\
\nu_{A+1}^{(i)}=-x_{A+1}^{(i)} + \omega \Delta_{A,A+1}^{(+)} \\
\mu_r= \{\mu_{A-1}^{(i)}, \mu_{A+1}^{(i)}\}
\end{array}
& \qquad\,\,
\begin{array}{l}
\Delta_{J+1,J} = \Delta_{J,J+1}^{(-)} \\
\mu_{A-1}^{(i)}=x_{A-1}^{(i)} +  \omega \Delta_{A-1,A}^{(+)} \\
\nu_{A-1}^{(i)}=-x_{A-1}^{(i)} + \omega \Delta_{A-1,A}^{(-)} \\
\nu_s= \{\nu_{A-1}^{(i)}, \nu_{A+1}^{(i)}\} 
\end{array}
\end{array}
\eea
which satisfy the superpotential contraint
\begin{equation}
\sum_{r=1}^{2N} \mu_r + \sum_{s=1}^{2N} \nu_s =  N \omega (\Delta_{A,A+1}^{(-)}+\Delta_{A,A+1}^{(+)}+\Delta_{A-1,A}^{(-)}+\Delta_{A-1,A}^{(+)})= 2 \omega  N  
\end{equation}
Here $r$ and $s$ are collective indices for elements of the respective sets.
By applying equation \eqref{Ucase} and 
fixing $k_A>0$ we obtain
\small
\bea
\mathcal{Z}_m&=& \frac{1}{\sqrt{(-\omega_1 \omega_2)^{N+k_A}} (N+k_A)!}
\int
\frac{\displaystyle\prod_{J=A\pm1} \prod_{i=1}^{N}\prod_{j=1}^{N+k_A} \Gamma_h\left(\omega-\nu_J^{(i)}-x_A^{(j)} \right) \, \Gamma_h\left(\omega-\mu_J^{(i)}+x_A^{(j)} \right)}
{\displaystyle \prod_{i<j}^{N+k_A} \Gamma_h \left(\pm \left( x_{A}^{(i)}-x_{A}^{(j)}\right)\right)} \nonumber \\
 \nonumber \\ &\,\,& 
 \prod_{i=1}^{N} c\left(-2 (k_A+k_J) {x_{J}^{(i)}}^2\right)\prod_{i=1}^{N+k_A}  c\left(2 k_A {x_A^{(i)}}^2\right) d x_{A}^{(i)} \times \prod_{r,s=1}^{2N} \Hyper{\mu_r+\nu_s} 
 \nonumber\\ 
 &&
  \nonumber\\ 
 &&
\zeta^{-k_A^2-2} \,\, c\left(
 k_A^2 \left(2 \omega ^2-1\right) +2 \, N\,  k_A\left(\omega ^2 \left(\Delta _{A-1,A}^2+\Delta _{A,A+1}^2-2\right)-1\right)
 \right) 
\label{eq:magpt}
\eea
\normalsize
The denominator can be interpreted as the 1-loop contribution from the vector superfield of the gauge group $U(N+k_A)$
(recall that the duality does not change the ranks of other factors).\footnote{In this case
 we choose all the ranks $N_J$ equal to $N$.  In more general situations, when fractional branes 
are considered in the electric theory, all the ranks can be different, and the duality preserves the partition
function as in this case.
Moreover, as explained in the appendix, we are restricting to $k_A>0$. For a generic $k_A$ the 
dual rank becomes $N+|k_A|$.}
  The numerator in the first term contains the contribution from 
 the (anti)bifundamental
fields: it is easy to see that  bifundamental fields are mapped to anti-bifundamental fields and viceversa,
as required by Seiberg duality. Moreover, we also obtain the offshell map between the scaling dimensions $\tilde\Delta$
of the dual fields and the electric ones
\bea  \label{dualcharges}
\widetilde \Delta_{A,A\pm1}= 1-\Delta_{A\pm1,A}
\qquad\qquad
\widetilde \Delta_{A\pm1,A}= 1-\Delta_{A,A\pm1}
\eea
The last factor in the second line of  \eqref{eq:magpt} gives the contribution from the new adjoint fields. Indeed, it can be
written in the form
\bea \label{mesons}
\prod_{r,s=1}^{2N} \Gamma_h(\mu_r+\nu_s)
&=&
\prod_{J=A\pm1} \prod_{i,j=1}^N
\Gamma_h\left( x_{J}^{(i)} - x_{J}^{(j)}+\omega \widetilde\Delta_{J,J}\right)  \times
\nonumber \\
&& \prod_{i,j=1}^N
\Gamma_h\left(  x_{A+1}^{(i)} -   x_{A-1}^{(j)} +\omega (\Delta_{A,A+1}^{(+)} + \Delta_{A-1,A}^{(-)})\right)
\nonumber \\
&&  \prod_{i,j=1}^N
\Gamma_h\left(  x_{A-1}^{(i)} -   x_{A+1}^{(j)} +\omega  (\Delta_{A,A+1}^{(-)} + \Delta_{A-1,A}^{(+)})\right)
\eea
where $\widetilde \Delta_{A\pm1,A\pm1} = \Delta_{A\pm1,A}+\Delta_{A,A\pm1}$ gives the $R$-charge
of the adjoint fields. 
On the field theory side the dual superpotential is
\bea
W = \dots &+&Y_{A,A-1} Y_{A-1,A-1} Y_{A-1,A}   + Y_{A-1,A}Y_{A,A+1} Y_{A+1,A-1} \nonumber \\&+&  Y_{A+1,A}Y_{A,A-1} Y_{A-1,A+1}
+Y_{A,A+1} Y_{A+1,A+1} Y_{A+1,A}
+\dots
\eea
and by integrating out the fields $Y_{A-1,A+1}$ and $Y_{A+1,A-1}$
it becomes
\bea
W = Y_{A,A-1} Y_{A-1,A-1} Y_{A-1,A} - Y_{A-1,A} Y_{A,A+1} Y_{A+1,A} Y_{A,A-1}+Y_{A,A+1} Y_{A+1,A+1} Y_{A+1,A} +\dots
\nonumber \\ 
\eea
where the  dual fields $Y_{A\pm1,A\pm 1}$ are related to the electric ones as
\begin{equation}
Y_{A\pm1,A\pm 1} = X_{A\pm 1,A}X_{A,A\pm1}
\end{equation}
Formula (\ref{mesons}) takes properly into account the contribution of the new mesons $Y_{A+1,A+1}$ and $Y_{A-1,A-1}$.
The contribution of the two extra mesons reduces to $1$ in (\ref{mesons}) after using the  reflection formula (\ref{refle}) 
and the superpotential 
constraint$\widetilde  \Delta_{A-1,A+1} +\widetilde  \Delta_{A+1,A-1}=2$.

We now check that the CS levels shift according to the discussion in section \ref{sec:models}.
For simplicity we gauge fix
the complexified Fayet-Iliopoulos (FI) term $\Delta_m$ to zero,
but the corresponding generalization is straightforward and
one can easily map the electric FI in the magnetic one as
$\Delta_m'$ = $\Delta_m'\left (\Delta_m,\Delta_{J,J+1}^{(\pm)}\right)$.
We stress that we can perform this gauge fixing choice without worrying about extremization with
respect to $\Delta_m$ because we consider $U(N)$ factors as opposed to $SU(N)$ ones. Below, when
we will consider orthogonal and symplectic gauge groups, the FI term will vanish even for simple group
factors because of invariance under charge conjugation.

Having fixed the FI term, the linear terms in the function $c$ in \eqref{eq:magpt} have to cancel out.
Recall that in a vector-like theory with vanishing FI term we also have
$\Delta_{J,J+1}^{(+)}= \Delta_{J,J+1}^{(-)}$  and $\sum \mu_r=\sum \nu_s$.
We only need these relations here, but they can be easily relaxed if one wishes to introduce a nontrivial
FI term in the model.
We obtain in (\ref{eq:magpt}) the shift of the levels $k_{A\pm1}$ by a factor of $k_A $ while  the level for the 
dualized group switches its sign.
Finally the last line in  (\ref{eq:magpt}) represents a pure
phase factor, which does not spoil the duality.

\subsubsection{Adding an adjoint field}

\begin{figure}[htbp]
\begin{center}
\includegraphics[width=10cm]{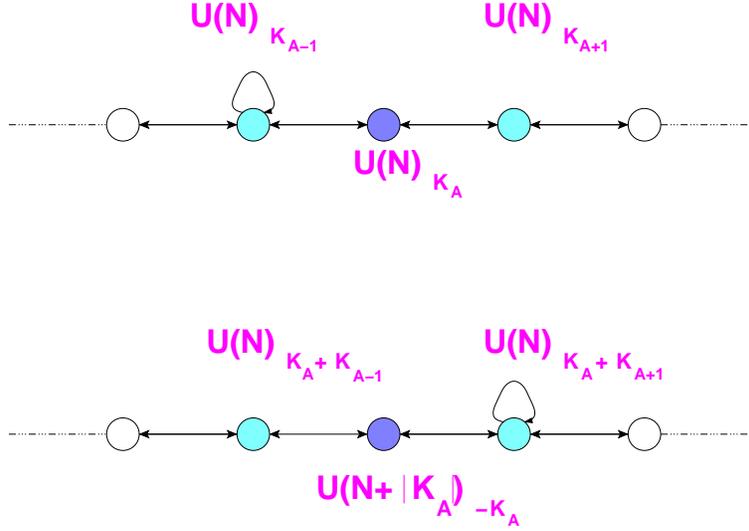}
\end{center}
\caption{Dual phases describing a stack of M$2$ branes probing Calabi-Yau fourfold with adjoint matter involved in the duality. }
\label{Figus}
\end{figure}

We now consider a slightly different model which also contains an adjoint field $X_{A-1,A-1}$ on the electric side.
The quiver for the dual phases is depicted in Figure \ref{Figus}.
%
%
%
%
%
%
The superpotential for the colored nodes of the quiver is
\bea \label{eq:Welecadj}
W_e =  \dots+X_{A-1,A-1} X_{A-1,A} X_{A,A-1} - X_{A-1,A} X_{A,A+1} X_{A+1,A} X_{A,A-1} +\dots
\eea
The dual superpotential is
\bea \label{eq:Welecadj2}
W_m =  \dots Y_{A-1,A} Y_{A,A+1} Y_{A+1,A} Y_{A,A-1} -Y_{A+1,A+1} Y_{A+1,A} Y_{A,A+1} +\dots
\eea
The relevant contribution to the electric partition function  on  the squashed sphere is:
\begin{eqnarray}
\label{eq:Zelecadj}
\mathcal{Z}_e = 
&&
\frac{1}{\sqrt{\omega_1^N\omega_2^N} N!} 
\int \frac{\displaystyle\prod_{J=A-1}^{A} \prod_{i,j=1}^{N} \Gamma_h\left(\pm \left( x_J^{(i)} -x_{J+1}^{(j)} \right) + \omega \Delta_{J,J+1}^{(\eta)} \right) }
{\displaystyle \prod_{J=A-1}^{A+1} \prod_{i<j}^{N} \Gamma_h \left(\pm \left( x_{J}^{(i)}-x_{J}^{(j)}\right)\right)}
 \\
&&
\prod_{i,j=1}^{N}
\Gamma_h\left( \pm \left( x_i^{(A-1)}-x_i^{(A-1)}\right)+ \omega \Delta_{A-1,A-1}
\right)
\prod_{J=A-1}^{A+1} \prod_{i=1}^{N} c\left(-2 k_J {x_J^{(i)}}^2\right) d x_{J}^{(i)}
\nonumber
\end{eqnarray}
The duality can be shown 
by following the same steps as in Subsection \ref{dualityUN}.
The only difference is that in (\ref{mesons}) there is an extra constraint 
$\Delta_{A-1,A-1} + \Delta_{A-1,A}+\Delta_{A,A-1}=2$.
This constraint sets the contribution of the meson $Y_{A-1,A-1}$ 
to $1$ in the dual partition function (in field theory it is integrated out)
because of (\ref{refle}).
\subsection{The first  class of orientilfolds: O$3$ planes}

In this section we study the duality on the first class of 
 orientifolded models introduced in section \ref{subsect:ori} and 
 match the partition function between different phases.
Recall that the relevant models are quiver field theories with alternating $"a"$
$SP(2N)_{k_i}$ and $"a"$ $SO(2N)_{2 k_i}$ nodes, with $\sum k_i=0$.
The superpotential is
\begin{equation}
W = \sum_{J=1}^{2 a-1}  \left(X_{J,J+1} \cdot X_{J+1,J+2}\right)^2
\end{equation}
where $X_{2a,2a+1}=X_{2a,1}$.
If $a>1$ there is always a field connecting two 
consecutive nodes labeled by   $J$ and $J+1$, and we assign to this field the charge
$\Delta_{J,J+1}$.\footnote{The case $a=1$
reduces to the models studied in \cite{Aharony:2008gk}.}
The superpotential imposes the constraint $\Delta_{J-1,J}+\Delta_{J,J+1}=1$.

\subsubsection{Duality on an $SP(2N)_k$ node}

We first study the duality on an $SP(2N)_{k_A}$ group.
Also in this case we refer to the quiver in Figure \ref{Figu}, but we erase the
arrows because the groups are real and there is no distinction between fundamental
and antifundamental representations.
The relevant contribution to the partition function for this model is
\begin{equation}
\mathcal{Z}_{SP(2N)_{k_A}} =
\int \frac{\displaystyle \prod_{J=A-1}^{A} \prod_{i,j=1}^N \Gamma_h\left(\pm x_J^{(i)}\pm x_{J+1}^{(j)} +\omega \Delta_{J,J+1}\right)
\prod_{J=A-1}^{A+1} \prod_{i=1}^{N} c\left(-4 k_{J} {x_{J}^{(i)}}^2\right) }
{ \displaystyle\prod_{1\leq i<j\leq N} \Gamma_h\left( \pm x_A^{(i)} \pm x_{A}^{(j)}\right) \prod_{i=1}^{N} \Gamma_h\left(\pm 2 x_A^{(i)}\right)}
\prod_{i=1}^{N}  dx_{A}^{(i)}
\end{equation}
where we used  the notation 
$\Gamma_h(x+a) \Gamma_h(- x+a) =\Gamma_h(\pm x+a) $.
In this case we define the $\mu_r$ variables as
\begin{equation}
\mu_{i,{A-1}}^{(\pm)} = \pm x_{A- 1}^{(i)} +\omega \Delta_{A- 1,A}
\quad , \quad
\mu_{i,{A+1}}^{(\pm)} = \pm x_{A+ 1}^{(i)} +\omega \Delta_{A,A+1}
\end{equation}
Since there are $4N$ different $\mu$ the index $r$ runs 
from $1$ to $4N$, such that
\begin{equation}
\mu_r = \{ \mu_{i,A-1}^{(+)},  \mu_{i,A-1}^{(-)}, \mu_{i,A+1}^{(+)}, \mu_{i,A+1}^{(-)}\}
\end{equation} 
where every $i$ runs from $1$ to $N$.
The dual gauge group is 
\begin{equation}
SO(2 N)_{2(k_{A-1}+k_{A})}
\times
SP(2(N+|k_A|-1))_{-k_A}
\times
SO(2 N)_{2(k_{A}+k_{A+1})}
\end{equation}
The dual superpotential is
\begin{equation} \label{dualSPspot}
W_m=
Y_{A\pm1,A\pm1} \cdot Y_{A\pm1,A} \cdot Y_{A,A\pm1}
-
\left(Y_{A-1,A}\cdot Y_{A,A+1}\right)^2
\end{equation}
The partition function of the dual gauge theory
corresponds to  the RHS
of (\ref{SPcase}) by fixing $k_A>0$. 
In this  case
we have
\begin{eqnarray} \label{dualitySPg}
&&
I_{N,2(1+k_A)a}^{N+k_A-1}(\mu)=
I_{N+k_A-1,2(1+k_A)b}^{N}(\omega-\mu)
\prod_{1\leq r<s\leq 4N} \Gamma_h\left(\mu_r+\mu_s\right)
\zeta^{(k_A-1)(1-2k_A)}\\
&&
c\left(  
\omega ^2 \left(2 k_A^2\!-\!k_A\! \left(\!3\!+\!4 N \!\left(\!\Delta _{A-1,A}^2\!+\!\Delta _{A,A+1}^2\!-\!1\!\right)\!\right)\!+\!1\right)
\!-\!4 k_A\left(\sum_{i=1}^{N}{x_{A-1}^{(i)}}^2\!+\!
\sum_{i=1}^{N}{x_{A+1}^{(i)}}^2\!\right)\!
\right)  \nonumber  
\end{eqnarray}
The case $k_A<0$ in the electric
 theory is studied by inverting  (\ref{SPcase}) as explained in  Appendix \ref{vdb section}. 
As expected the rank of the dual groups is
$\tilde N = N+|k_A|-1$.

It is straightforward to see from the first term in the RHS of  
(\ref{dualitySPg}) that  the electric $R$-charge
of a bifundamental connecting  a pair of nodes   
in the electric theory 
is related in the magnetic theory
to the 
$R$-charge
of a bifundamental connecting the same pair of nodes
 nodes through 
$\widetilde \Delta_{i,j}=1-\Delta_{i,j}$.

The second term in the RHS of (\ref{dualitySPg}) can be expanded in terms of 
$\mu_r$ and it becomes
\begin{eqnarray}
\prod_{1\leq r<s\leq 4N} \Gamma_h\left(\mu_r+\mu_s\right)&=&
 \\
&=&
\prod_{1\leq i<j \leq N}
\Gamma_h\left(
\pm x^{(i)}_{A-1} \pm x^{(j)}_{A-1} + 2 \omega \Delta_{A-1,A}\right)
\times
{\Gamma_h}^N\left( 2 \omega \Delta_{A-1,A}\right)
\nonumber \\
&\times&
\prod_{1\leq i<j \leq N}
\Gamma_h\left(
\pm x^{(i)}_{A+1} \pm x^{(j)}_{A+1} + 2 \omega \Delta_{A,A+1}\right)
\times  {\Gamma_h}^N
\left( 2 \omega \Delta_{A,A+1}\right)
\nonumber \\
&\times&
\,\,\,\,\,\prod_{i,j=1}^N\,\,\,\,
\Gamma_h\left(
\pm x^{(i)}_{A+1} \pm x^{(j)}_{A-1} +  \omega \Delta_{A-1,A}+\omega \Delta_{A,A+1}\right)
\nonumber \\
\end{eqnarray}
The first two terms are the mesons of the dual theory while the 
last one evaluates to $1$  because  of the superpotential
constraint on the
R-charges.

We conclude the proof of the duality with the analysis of the 
CS contributions to the partition function.
The CS of the dual $SP$ group switches from $k_A$ to $-k_A$,
because the dual theory is a ``$b$" integral (see Appendix \ref{vdb section} for details).
The CS of the  $SO$ groups transform in (\ref{dualitySPg})
as $2k_{A\pm1}\rightarrow 2k_{A}+2k_{A\pm1}$, 
as expected.

Similar to the case of  unitary theories,  (\ref{dualitySPg})  also has 
which we ignore.

\subsubsection{Duality on an $SO(2N)_{2k}$ node}

In the O$3^{\pm}$ orientifolded quiver one can also 
dualize an $SO(2N)_{2k_A}$ node.
The dual gauge group is
\begin{equation}
SP(2N)_{k_{A-1}+k_A} \times SO(2(N+|k|+1))_{-2 k_A} \times SP(2N)_{k_{A}+k_{A+1}}
\end{equation}
and the superpotential is again (\ref{dualSPspot}) with the proper products.
The relevant contribution to the  partition function of the electric theory is 
\begin{equation}
\mathcal{Z}_{SO(2 N)_{k_A}} =
\int \frac{\displaystyle\prod_{J=A-1}^{A} \prod_{i,j=1}^{N} \Gamma_h\left(\pm x_J^{(i)}\pm x_{J+1}^{(j)} +\omega \Delta_{J,J+1}\right)
\prod_{J=A-1}^{A+1}
\prod_{i=1}^{N} c\left(-4 k_{J} {x_{J}^{(i)}}^2
\right)}
{\displaystyle \prod_{1\leq i<j\leq N} \Gamma_h\left( \pm x_A^{(i)}  \pm x_{A}^{(j)}\right)}
\prod_{i=1}^{N}  dx_{A}^{(i)}
\end{equation}
As in \cite{Willett:2011gp,Benini:2011mf} the measure of the  $SO(2N)$  gauge group 
can be converted into that of an $SP(2N)$ group 
by applying the relation (\ref{usefullater}) and inserting in the partition function the contribution
\begin{equation} \label{trick}
1=\frac{\displaystyle\prod_{i=1}^{N} \prod_{\alpha=1}^{4} \Gamma_h(\pm x_{A}^{(i)}+\rho_\alpha)}{\Gamma_h({\pm 2 x_{A}^{(i)}})}
\end{equation}
where $\rho_{\alpha} = \left(0,\frac{\omega_1}{2},\frac{\omega_2}{2},\omega\right)$.
The  $\mu$ vector becomes
\begin{equation}
\mu_r = \{ \mu_{i,A-1}^{(+)},  \mu_{i,A-1}^{(-)}, \mu_{i,A+1}^{(+)}, \mu_{i,A+1}^{(-)},\rho_{\alpha}\}
\end{equation} 
where $r=1,\dots,4N+4$ and
\begin{equation}
\mu_{i,{A-1}}^{(\pm)} = \pm x_{A- 1}^{(i)} +\omega \Delta_{A- 1,A}
\quad , \quad
\mu_{i,{A+1}}^{(\pm)} = \pm x_{A+ 1}^{(i)} +\omega \Delta_{A,A+1}
\end{equation}
By applying (\ref{SPcase}) with $k_A>0$ we have
\begin{eqnarray} \label{dualitySO}
I_{N,2(1+k_A)_a}^{N+k_A+1}(\mu)&=&
I_{N+k_A+1,2(1+k_A)_b}^{N}(\omega-\mu)
\prod_{1\leq r<s\leq 4N+4} \Gamma_h\left(\mu_r+\mu_s\right)
\zeta^{(k_A-1)(1-2k_A)} \nonumber \\
&\times&
c\left(
4 k_A \left(\sum _{i=1}^N {x_{A-1}^{(i)}}^2+\sum _{i=1}^N {x_{A+1}^{(i)}}^2\right)
-\frac{1}{2} k_A \left(\omega _1{}^2+\omega _1{}^2\right)
\right) \\
&\times &
c\left(
\omega ^2\left(2 k_A{}^2+k_A \left(3-4 N \left(\Delta _{A-1,A}{}^2+\Delta _{A,A+1}{}^2-1\right)\right)+1\right)
\right) \nonumber 
\end{eqnarray}
The case $k_A<0$ in the electric 
theory is studied by inverting (\ref{SPcase}).
 As  expected the rank of the dual groups is
$\tilde N = N+|k_A|+1$.
Observe that it fits with the proposal of \cite{Kapustin:2011vz}, $SO(\tilde N_c) = SO( N_f+|K|-2-N_c)$.
Indeed in our case $N_c=2 N $, $K=2 k_A$, $N_f=4 N+4$ and $\tilde N_c = 2(N+|k_A|+1)$.

The RHS of (\ref{dualitySO}) corresponds to the
partition function of the dual theory.
By using the relation (\ref{trick})
the extra terms in the measure arising in (\ref{dualitySO}) become
\begin{equation} \label{tricktransformed}
\frac{\displaystyle\prod_{i=1}^{N+|k_a|+1} \prod_{\alpha=1}^{4} \Gamma_h(\omega \pm x_{A}^{(i)}-\rho_\alpha)}{\Gamma_h({\pm 2 x_{A}^{(i)}})}=1
\end{equation}
thus giving us the 
measure of the $SO(2(N+|k_A|+1))$
dual gauge group.

Upon expanding the $\Gamma_h(\mu_r+\mu_s)$
term in the RHS of (\ref{dualitySO})
we find
\small
\begin{eqnarray} \label{mesonsSP}
\prod_{1\leq r<s\leq 4N+4} \hspace{-.2cm}\Gamma_h\left(\mu_r+\mu_s\right) 
=
\prod_{1\leq r<s\leq 4N}  \hspace{-.3cm}\Gamma_h\left(\mu_r+\mu_s\right)
 \hspace{-.3cm}
\prod_{{\tiny
\begin{array}{c}
1\leq r\leq 4N\\
4N<s<4N+4
\end{array}}}  \hspace{-.5cm}\Gamma_h\left(\mu_r+\mu_s\right)
\prod_{4N< r<s\leq 4N+4}  \hspace{-.5cm}\Gamma_h\left(\mu_r+\mu_s\right)
 \nonumber\\
\end{eqnarray}
\normalsize
By combining the first two products we obtain
\begin{equation} 
\prod_{i,j=1}^N \Gamma_h(\pm x_{A+1}^{(i)} \pm x_{A+1}^{(j)} +2 \omega \Delta_{A,A+1})
\times
\prod_{i,j=1}^N \Gamma_h(\pm x_{A-1}^{(i)} \pm x_{A-1}^{(j)} +2 \omega \Delta_{A-1,A})
\end{equation}
which represent the massless mesons of dual theory (they are the adjoints of
neighbouring  SP$(2N)$).
The extra contributions from the first two  terms in the RHS of (\ref{mesonsSP})
correspond to the massive mesons and evaluate to $1$.
The last term in  the product in (\ref{mesonsSP}) is 
\begin{equation}
\Gamma_h\left(\frac{\omega_1}{2}\right) 
\Gamma_h\left(\frac{\omega_2}{2}\right) 
\Gamma_h(\omega)^2
\Gamma_h\left(\omega+\frac{\omega_1}{2}\right)  
\Gamma_h\left(\omega+\frac{\omega_2}{2}\right) 
=1
\end{equation}
bacause of (\ref{otherid}).
The rest of the terms in  (\ref{dualitySO}) give the right transformation on the $CS$ levels and an
extra phase as usual.

\subsection{The second class of orientifolds: duality on  $SO(2N+1)_k$ }

If we consider  $\widetilde{\text{O}3}^{\pm}$ orientifold planes, 
the gauge groups of the necklace quiver 
involve
$SO(2N+1)$ factors instead of $SO(2N)$.
We are interested in studying the duality on these nodes.
The relevant contribution to the partition function is
\small
\begin{eqnarray} \label{SO2n+1}
\mathcal{Z}_{SO(2N+1)_{k_A}} &=& 
\int \frac{\displaystyle \prod_{J=A-1}^{A} \prod_{i,j=1}^{N} \Gamma_h\left(\pm x_J^{(i)}\pm x_{J+1}^{(j)} +\omega \Delta_{J,J+1}\right)
\prod_{J=A-1}^{A+1} \prod_{i=1}^{N} c\left(-4 k_{J} {x_{J}^{(i)}}^2\right) 
}
{\displaystyle \prod_{1\leq i<j \leq N} \Gamma_h\left( \pm x_A^{(i)} \pm x_{A}^{(j)}\right) \prod_{i=1}^{N}  \Gamma_h\left(\pm x_{A}^{(i)}\right)}
\nonumber
\\
&\times&
\prod_{i=1}^{N}
\Gamma_h\left(\pm x_{A-1 }^{(i)}+\omega \Delta_{A-1,A}\right)
\Gamma_h\left(\pm x_{A+1 }^{(i)}+\omega \Delta_{A,A+1}\right)
 dx_{A}^{(i)}
\end{eqnarray}
\normalsize
The measure of the $SO(2N+1)$ group can  be converted into the one of an $SP(2N)$ 
group by  
applying (\ref{usefullater}). We find
\begin{equation}\label{barbatrucco}
\frac{1}{\Gamma_h\left({\pm x_{A}^{(i)}}\right)}
 = \frac{\Gamma_h\left(\pm x_{A}^{(i)}+\frac{\omega_1}{2}\right)
\Gamma_h\left(\pm x_{A}^{(i)}+\frac{\omega_2}{2}\right)}
{\Gamma_h\left( \pm 2 x_{A}^{(i)}\right)}
\end{equation}
In this case the $\mu$ vector is  $4N+2$ dimensionful. The first $4N$ 
elements are the same of the previous orthogonal case
while the extra two are $\frac{\omega_1}{2}$ and $\frac{\omega_2}{2}$.

The partition function of the dual theory is obtained by applying 
(\ref{SPcase}) to (\ref{SO2n+1}).
By fixing $k_A>0$ we have
\begin{eqnarray} \label{dualitySO2n+1}
I_{N,2(1+k_A)_a}^{N+k_A}(\mu)&=&
I_{N+k_A,2(1+k_A)_b}^{N}(\omega-\mu)
\prod_{1\leq r<s\leq 4N+2} \Gamma_h\left(\mu_r+\mu_s\right)
\zeta^{(k_A+1)(1-2k_A)} 
\nonumber \\
&\times&
c\left(
-4 k_A \left(
\sum _{i=1}^N {x_{A-1}^{(i)}}^2+\sum _{i=1}^N {x_{A+1}^{(i)}}^2\right)
-\frac{1}{2} k_A \left(\omega _1^2+\omega _2^2\right)\right)
 \\
&\times&
c\left(
\omega ^2 k_A \left(2 k_A+1+4 N-4 N \left(\Delta _{A-1,A}^2+\Delta _{A,A+1}^2\right)\right)
\right)\nonumber
\end{eqnarray}
The case $k_A<0$ in the electric theory is studied by inverting (\ref{SPcase}).
As  expected the rank of the dual group is
$\tilde N = N+|k_A|$.
Observe that it fits with the proposal of \cite{Kapustin:2011vz}, $SO(\tilde N_c) = SO( N_f+|K|-2-N_c)$.
Indeed in our case $N_c=2 N +1$, $K=2 k_A$, $N_f=4 N+2$ and $\tilde N_c = 2(N+|k_A|)+1$.

As before we can transform the measure back to $SO(2( N+|k_A|) +1)$ 
by applying (\ref{barbatrucco}).
Then we study the mesons:
we have to expand  the product
\begin{equation}
\prod_{1\leq r<s\leq 4N+2} \Gamma_h\left(\mu_r+\mu_s\right) 
\times
\prod_{1\leq r\leq 4N} 
\Gamma_h\left(\mu_r \right)
\end{equation}
where the first term come from (\ref{dualitySO2n+1}) and the second one from (\ref{SO2n+1}).
It is not difficult to recognize the contribution of the dual 
mesons predicted by the duality.
The term $1\leq r<s\leq4N$ gives 
\begin{equation}  \label{mesonsprimo}
\prod_{1 \leq i< j \leq N} \Gamma_h(\pm x_{A\pm1}^{(i)} \pm x_{A\pm1}^{(j)} +2 \omega \Delta_{A,A\pm1})
\end{equation}
The extra  contributions come from 
\begin{equation}
\prod_{1\leq r<\leq 4N, 4N<s \leq 4N+2} \Gamma_h\left(\mu_r+\mu_s\right) 
\times
\prod_{1\leq r\leq 4N} 
\Gamma_h\left(\mu_r \right)
\end{equation}
Explicitly we have
\begin{eqnarray}
&&\prod_{i=1}^{N} \left(\prod_{\alpha=1,2}
\Gamma_h\left(\pm x_{A\pm1}^{(i)} + \omega \Delta_{A\pm1} + \frac{\omega_\alpha}{2}\right)
\right)
\times
\Gamma_h\left(\pm x_{A\pm1}^{(i)} + \omega \Delta_{A\pm1}\right)=\nonumber
\\
=
&&
\frac{\Gamma_h\left(\pm 2 x_{A\pm1}+\omega \Delta_{A\pm1,A}\right)}{\Gamma_h\left(\pm  x_{A\pm1}+\omega \Delta_{A\pm1,A}+\omega\right)}
\end{eqnarray}
The numerator  in this expression replaces $i<j$ with $i \leq j$ in  (\ref{mesonsprimo})
while the denominator can be transformed as
\begin{equation}
\frac{1}{\Gamma_h\left(\pm  x_{A\pm1}+\omega \Delta_{A\pm1,A}+\omega\right)}
=
\Gamma_h\left(\pm  x_{A\pm1}+\omega (1-\Delta_{A\pm1,A})\right)
\end{equation}
which corresponds to the dual of the second line of (\ref{SO2n+1}).

\section{Duality and free theories: some exact results} \label{sec:free}
 
Three dimensional dualities are not only important for theories with an AdS dual 
but  also for more general SCFTs.
For example in \cite{Jafferis:2011ns} a new duality was proposed between an $SU(2)_1$
CS theory with an adjoint and no superpotential and a free theory.
This duality was further studied in \cite{Kapustin:2011vz}, 
in which an interacting CS matter theory without superpotential
is dual to a free theory.
While many checks have been performed by expanding  the superconformal index, 
and comparing the expansions on both sides of the dualities, a full understanding of
the matching of the partition function is  still missing.
Here we show the matching between the partition functions analytically.

The models considered in this section do not suffer from accidental symmetries.
In every case the partition function matrix integral of the electric interacting theory can
be worked out exactly and the extremization of the result sets the $R$-charges of the magnetic
fields to the canonical value without any modification.
In general, the naive extremization does not give this result, because the $R$-symmetry mixes
with accidental symmetries. We comment on the latter cases in the next section.

A technical comment is in order.
In the following we need some relations involving the integrals dubbed as $\textit{JI}$ in appendix
\ref{vdb section}. We take them from \cite{fvdb} and mention them in the text when necessary.

\subsection{$SU(2)_1$ theory with an adjoint field}

The first example is an $SU(2)_1$ CS theory
with an adjoint, studied in \cite{Jafferis:2011ns}.
The authors proposed a general formula for the partition 
function in this case
but they did not prove this formula analytically.
Here we use the results of \cite{fvdb} to show the agreement.
The partition function on the round sphere is  
\begin{eqnarray}\label{eq:Jafferis-Yin}
Z_{SU(2)_1}(\Delta)&=&\int dx 
\sinh^2\left( 2 \pi x \right)  e^{2 \pi i x^2}
e^{l(1-\Delta) + l(1-\Delta+2 i x)+l(1-\Delta-2 i x)} \\
&=&
\frac{1}{4}
\int d\epsilon \int dx_1 dx_2 \left(
-4 \sin (\frac{\pi (x_1-x_2)}{i}) \sin(\frac{\pi (x_1-x_2)}{i})
\right) \nonumber \\
&& 
e^{\pi i (x_1^2+x_2^2)}
e^{l(1+ i \tau) + l(1+i \tau+i (x_1-x_2))+l(1+i \tau - i (x_1-x_2))}
e^{2 \pi i (x_1+x_2) \epsilon }\nonumber 
\end{eqnarray}
where we used the relation
\begin{equation}
\int d \, x_1 d \, x_2 \delta(x_1+x_2) = \int d \epsilon \int d \, x_1 d \, x_2 e^{2 \pi i (x_1+x_2)\epsilon}
\end{equation}
and we set  $\tau= i \Delta$, where $\Delta$ represents the $R$-charge of the adjoint field. 
In terms of the hyperbolic functions the partition function becomes
\begin{eqnarray}
Z_{SU(2)_1}(\Delta)=
&&
-\frac{1}{4 \Gamma_h'(\tau)}
\int d \epsilon \Bigg( \frac{\Gamma_h'(\tau)^2}{2} \int d\,x_1 d\,x_2 
\frac{\Gamma_h'(\tau\pm(x_1-x_2))}{\Gamma_h'(\pm(x_1-x_2)}
\nonumber \\
&& \times
c'(2 \lambda (x_1+x_2)-2 (x_1^2+x_2^2))
\Bigg)
\end{eqnarray}
where  $\lambda=-2 \epsilon$. We used the notations $\Gamma_h'$ and $c'$ to specify that 
we are considering $\omega_1=\omega_2=i$, i.e. this is the partition function on the three sphere.
More generally the  formula inside the parenthesis 
can be associated to the partition function on the squashed three sphere,
and the resulting integral has been computed in  \cite{fvdb}. Here we quote the result
\begin{eqnarray}
\label{formulaadjsu}
&&
\frac{\Gamma_h(\tau)^n}{\sqrt{\left(-\omega_1\omega_2\right)^n} n!}
\int \prod_{1<\leq i\leq j\leq n} 
\frac{\Gamma_h(\tau \pm (x_i - x_j)}{\Gamma_h(\pm (x_i - x_j))}
\prod_{j=1}^{n} c(2 \lambda x_j - 2 x_j^2) dx_j =\nonumber \\
= &&
\zeta^{-3n}
\prod_{j=1}^{n} \Gamma_h \left( j \tau\right) 
c\left(\frac{n}{2} \left(2 \omega^2+\lambda^2+2(n-1) \tau \omega +\frac{1}{3}(n-1)(2n-1)\tau^2\right)\right)
\end{eqnarray}
By reducing on the three sphere, fixing $n=2$  and applying (\ref{formulaadjsu})  we have
\begin{equation} \label{afterVdB}
Z_{SU(2)_1} = 
-\frac{ 1}{4 \Gamma_h'(\tau)}
\int d \epsilon \,\,\,
{\zeta'}^{-6} \Gamma_h'(\tau) \Gamma_h'(2\tau)
c'\left(-2+\lambda^2 +2 i \tau +\tau^2
\right)
\end{equation}
If we substitue $\zeta' = e^{i \pi/12}$ and $\tau=i\Delta$
in (\ref{afterVdB}) and  perform the gaussian  integration
\begin{equation}
\int_{-\infty}^{\infty}
d \, \lambda e^{-\frac{i \pi \lambda^2}{2}} = \frac{2}{\sqrt 2} e^{-\frac{i \pi}{4}}
\end{equation}
the final expression becomes 
\begin{equation}
Z_{SU(2)_1} = 
\frac{1}{2 \sqrt{2}} e^{l(1-2 \Delta)} e^{\frac{i \pi}{2} (1+\Delta)^2 -\frac{i \pi}{4}}
\end{equation}
which coincides with the one proposed by \cite{Jafferis:2011ns}.

\subsection{SO$(4)_1$ with the adjoint field} \label{secSOtriv}

As discussed in \cite{Kapustin:2011vz}, the $SO(4)_1$ 
with an adjoint reduces to two copies of the \cite{Jafferis:2011ns} duality, 
because $SO(4) \simeq SU(2)\times SU(2)$.
Here we show that the partition function can be exactly computed in the $SO$ 
cases by using the results of \cite{fvdb} and reproduce the $SO(4)_1$ case explicitly.
In this case we need the relation 
\begin{eqnarray} \label{vdbformulaSO}
&& \frac{\Gamma_h(\tau)^n}{\sqrt{\left(\omega_1\omega_2\right)^n }2^n n }
\int 
\frac{\displaystyle \prod_{1\leq i<j \leq n} \Gamma_h(\tau \pm x_i \pm x_j)\prod_{j=1}^n \prod_{r=1}^{3}\Gamma_h(\mu_r\pm x_j)}
{\displaystyle \prod_{1\leq i<j \leq n} \Gamma_h(\pm x_i \pm x_j) \prod_{j=1}^n \Gamma_h(\pm 2 x_j)} \prod_{j=1}^{n} c(-2 x_j^2) d x_j=
\nonumber 
\\
=&&
\prod_{j=1}^{n} \Gamma_h( j \tau)
\prod_{1\leq r<s \leq 3} \Gamma_h(j \tau +\mu_r +\mu_s)
\\
\times &&
c\left(n\left(2(\mu_0 \mu_1+\mu_1\mu_2+\mu_2\mu_0)+2(n-1)\tau\sum_{r=1}^{3}\mu_r +\frac{1}{3}(n-1)(4 n-5) \tau^2\right)\right)
\nonumber
\end{eqnarray}
This equation can be applied to the SO$(2 n)_1$ case after we identify
\begin{equation}
\mu_1=0\quad,\quad\mu_2=\frac{\omega_1}{2}\quad,\quad\mu_3=\frac{\omega_2}{2}
\end{equation}
because, by applying (\ref{refle}) and (\ref{usefullater}),
we have
\begin{equation} \label{subst2in1}
\frac{\Gamma_h\left(\pm x\right) \Gamma_h\left(\pm x+\frac{\omega_1}{2}\right) \Gamma_h\left(\pm x+\frac{\omega_2}{2}\right)}
{\Gamma_h(\pm 2 x) }
=1
\end{equation}
Formula (\ref{vdbformulaSO}) then reduces to the partition function of a SO$(2 n)_1$ theory with a 
field in the adjoint representation.
 If we reduce to the case $n=2$ and fix $\tau=i \Delta$ and $\omega_1=\omega_2=i$ the partition function
 on the three sphere  is
\begin{equation}
e^{2 l (1-2 \Delta)} e^{2 \pi i \left(\Delta+\frac{1}{2}\right)^2-\frac{3}{2} \pi i }
\end{equation}
which reduces to two copies of $SU(2)_1$ theories with the adjoint and differs from that case just by a phase factor.

\subsection{$SP(4)_{2}$ with an absolutely antisymmetric field}

In the case of  symplectic groups  also there are exact relations in \cite{fvdb} that can be applied to obtain a CS matter theory
dual to a free theory.

Here we study the irreducible  absolutely  antisymmetric representation, described by the Dynkin label $\vec{s}=(1,1,0\dots,0)$
(see Appendix \ref{app:A} for details).
By using the Schur polynomial in the appendix the character of this irreducible representation is
\begin{equation}
\chi_{(1,1,0,\dots,0)}=\sum_{i\neq j} \left( z^i z^j+z^i z^{-j}+z^{-i}z^{j}+z^{-i} z^{-j}\right)+N-1
\end{equation}
In this case we need the equality \cite{fvdb}
\begin{eqnarray}
&& \frac{\displaystyle \Gamma_h(\tau)^{n-1}}{\sqrt{\left(-\omega_1 \omega_2\right)^n} 2^n n!}
\int
\prod_{1\leq i<j\leq n}
\frac{\Gamma_{h}(\tau\pm x_i \pm x_j)}{\Gamma_h(\pm x_i \pm x_j)}
\prod_{j=1}^{n}
\frac{1}{\Gamma_h(\pm 2 x_j)}
c(-8 x_j^2) d x_j = \nonumber \\
= &&
\zeta^{-3n} \prod_{j=2}^{n} \Gamma_h(j \tau) 
\,c\left(n\left(3 \omega^2+3(n-1)\tau\omega+\frac
{1}{6}(n-1)(2n-7)\tau^2\right)\right)
\end{eqnarray}
For $n=2$ it represents the partition function for 
a
$SP(4)_2$ gauge theory with an absolutely  antisymmetric 
two index tensor. By fixing $\tau=i \Delta$ and $\omega_1=\omega_2=i$ the
partition function on the three sphere becomes 
\begin{equation} \label{PFSP}
\mathcal{Z}_{SP(4)_{2}}= 
e^{l(1-2 \Delta)}  e^{-\frac{i \pi}{2} ( \Delta-3)^2 }
\end{equation}
This relation suggests that this theory is dual to a free theory with a singlet.

A similar duality appeared in \cite{Kapustin:2011vz}, however 
the antisymmetric representation considered there 
was not irreducible and contained another singlet.
This extra singlet adds a
$\Gamma_h(\tau)$ factor on both sides
of (\ref{PFSP}), leaving the equality unchanged.
In this case the theory is dual to a theory with two singlets with charges 
$\Delta$ and $2 \Delta$. 
This case contains accidental symmetries which
mix wit the $R$-symmetry and need to be properly accounted in the extremization
of the partition function.
We will comment on this issue in  section \ref{sec:acc}.

\subsubsection{The superconformal index}

The superconformal index is a Witten like index which counts over the protected 
BPS states of the theory. The index for three dimensional theories with $\mathcal{N}\geq 3$
SUSY
was first proposed in \cite{Bhattacharya:2008bja}   by localizing the theory on on $S^2 \times S^1$.
The expression for the index is given by
\begin{equation}
\mathcal{I}(x,y_i) = \text{Tr} (-1)^F x^{E+j_3} \prod_i y_i^{F_i}
\end{equation}
where $F$ is the fermion number, E is the energy,
$j_3$ is the third component of the $SU(2)$ rotational symmetry
in the superconformal group.
This index was refined to include the monopole  contributions  in \cite{Kim:2009wb}.
For theories with $\mathcal{N}=2$ supersymmetry the $R$-charge is not constrained 
to be canonical anymore, and the index for a generic $R$-charge assignment
was found in \cite{Imamura:2011su}.
It is important to observe that, in the general case, dual theories share the same 
index only after the contribution from the monopole sectors is included. In some 
cases  the index matches sector by sector but in general one has to sum over all the sectors.
For example if an interacting theory  is dual to a free theory one has to necessarily include the 
monopole corrections before matching the indices.

Here we consider the index of the $SP(4)_2$ CS theory with one matter field in the
absolutely antisymmetric representation and $R$-charge $\Delta$.
After including the  contribution from monopoles with GNO charge  $(1,0)$
the superconformal index is\footnote{GNO charges are
 quantum numbers labeling the different monopole sectors of the theory \cite{Goddard:1976qe} .
In the $SP(4)$ case the GNO charge of a sector carrying $m$ unit of magnetic  flux is $(m,0)$.}
\begin{eqnarray}
\mathcal{I}&=&  
\left(1-x^2+x^{2 \Delta }+x^{4 \Delta }+x^{6 \Delta }+x^{8 \Delta }+x^{10 \Delta }+x^{12 \Delta }+x^{14 \Delta } + \dots\right){}_{(0,0)} \\
&+&\left(-x^{2-2 \Delta }-x^{4-2 \Delta }+\dots\right){}_{(1,0)} +\dots \nonumber \\
&=&
1-x^2-x^{2-2 \Delta }-x^{4-2 \Delta }+x^{2 \Delta }+x^{4 \Delta }+x^{6 \Delta }+x^{8 \Delta }+x^{10 \Delta }+x^{12 \Delta }+x^{14 \Delta }+\dots
\nonumber
\end{eqnarray}
This coincides with the index of a free multiplet with $R$-charge $2 \Delta$, corroborating the duality proposed 
above.

\section{Comments on accidental symmetries}
\label{sec:acc}
 
In this section we briefly comment on a proposal to deal with accidental symmetries in three-dimensional field theories.
We will adapt to the three-dimensional case a similar prescription used in the four-dimensional $a$-maximization
\cite{Kutasov:2003iy},
with the respective physical meaning \cite{Barnes:2004jj}, which also allows for an extension away from the fixed points
\cite{Amariti:2011xp} based on the four dimensional analogy \cite{Kutasov:2003ux}. For a preliminary discussion, see \cite{Morita:2011cs}.

Any time the fixed point scaling dimension of a scalar gauge invariant operator drops below the $d$-dimensional unitary
bound $\Delta \ge (d-2)/2$, this signals that the UV description that we are using to extract information about the IR physics
is no longer valid, because the theory enjoys new "accidental" symmetries which are not manifest in the UV description.
The new symmetries are generated by the gauge invariant operators, which decouple from the rest of the theory in the
IR: they retain their canonical scaling dimensions and describe free fields. In these cases we need to modify the UV description
in a suitable way, which we describe in the following.

Consider a model where $m$ gauge invariant operators ${\cal O}_i$, $i=1,\ldots,m$ hit the unitary bound,
and consider coupling to that theory $m$ sources $L_i$ and $m$ gauge invariant operators $M_i$ through the
superpotential
\bea
\Delta W = L^i \left( {\cal O}_i + \lambda M_i \right)
\label{eq:DW}
\eea
where $\lambda$ is small in the UV. The operators ${\cal O}_i$ are, in general, not related to each other,
and so are the $L$'s and the $M$'s.
Imposing the condition $R(L_i)+R({\cal O}_i)=2$ we see that when $R({\cal O}_i)>(d-2)/(d-1)$
the last term is indeed relevant and makes the fields $L$ and $M$ massive.\footnote{Recall that in a superconformal
field theory the $R$-charge and the scaling dimension are related by $R=2\Delta/(d-1)$, where the
superpotential has $R$-charge $2$.}
Once they are integrated out, we
obtain the IR superconformal theory we started with, and no physical quantity has
changed.\footnote{This is a physical requirement on any physical quantity that depends on the
exact superconformal $R$-charges: the contribution from massive fields has to cancel out.\label{fn:massive}}
On the other hand,
when $R({\cal O}_i)<(d-2)/(d-1)$, the $L M$ coupling is irrelevant and the $M$'s are free decoupled fields in the IR.

In the case of three-dimensional field theories, where $R=\Delta$, a free field
contributes a factor $\exp\left( \ell(1/2) \right)=2^{-1/2}$ to the partition function, or equivalently a term $\log(2)/2$
to the free energy. The $R$-charge of the $L$'s is fixed by the first term in \eqref{eq:DW},
 and their contribution to the partition
function is $\exp\left(m\ell(-1+\Delta({\cal O})\right)$. Summing everything up, we obtain
\bea
F = F_0 +  \left(m  \frac{\log(2)}{2} + \sum_{i=1}^{m} \ell\left( 1-\Delta({\cal O}_i) \right) \right)
\label{eq:modifiedF}
\eea
where we also used $\ell(1-\Delta)=-\ell(-1+\Delta)$ for $0<\Delta<2$, which is always the case
in any sensible theory (see footnote \ref{fn:massive}). Equation
(\ref{eq:modifiedF}) has a very clear interpretation: along the RG flow, the $R$-charges as a
function of the RG scale are given by the Lagrange multiplier technique \cite{Amariti:2011xp};
when a gauge invariant operator hits the unitary bound,
one subtracts its contribution to the free energy and adds the contribution of the same number
of free fields. Because both the correction term to (\ref{eq:modifiedF}) and its first derivative vanish
at the free field point $\Delta({\cal O}_i)=1/2$, all the $R$-charges and the free energy itself are continuous
and differentiable functions of the RG scale.

\subsection{Accidental symmetries  in the duality with free theories}

In section \ref{sec:free} we focused on theories with a free magnetic dual whose
partition function can be exactly and consistently computed by localization and extremization
without any further modification. We now apply the discussion in the previous subsection
and show how the computation of the exact superconformal $R$-charge can be
consistently worked out even when the infrared theory enjoys accidental symmetries.
This provides new and stronger checks of three-dimensional dualities.

We start by describing an example in some detail, in which 
 the dual gauge group vanishes and the magnetic theory only contains
a tower of non-interacting singlets with naive $R$-charges different from the canonical ones.
The simplest electric theory of this kind has $U(N_c)_1$ gauge group and contains one adjoint $X$
with a vanishing superpotential \cite{Kapustin:2011vz}. 

The partition function of the $U(N_c)_1$ theory with one adjoint $X$ can be exactly computed \cite{fvdb}
\bea
\mathcal{Z}_{U(N_c)_1,X}= e^{\frac{i \pi }{12} N \left(3+6\Delta \left( N_c-1\right)+\Delta ^2 \left(2 N_c^2-3 N_c+1\right)\right)}  \times \prod_{j=1}^{N_c} e^{l(1- j \Delta)}
\label{PFUN}
\eea
where $j \Delta$ is the $R$-charge of Tr$X^j$. 
Notice that the $U(1)\subset U(N_c)$ decouples and 
Tr$X$ is a free field. However, for the sake of  uniform treatment,
we keep its $R$-charge to be $\Delta$ instead of $1/2$.

The naive $R$ charges of the $N_c$ free fields 
of the magnetic theory, given by $u_j=$Tr$X^j$,
are obtained by extremizing \eqref{PFUN}, 
which boils down to
solving the equation
\begin{equation}
\frac{\partial \log\left|\mathcal{Z}_{U(N_c)_1,X} (\Delta,N_c)\right|}{\partial \Delta}
=
\sum_{j=1}^{N_c}j \pi (1-j \Delta ) \cot(\pi(1-j \Delta))
=0
\label{eq:extrpart}
\end{equation}
The solution  is
\bea
\Delta = \frac{1}{N_c+1}
\label{eq:delta1}
\eea
Proving that \eqref{eq:delta1} solves \eqref{eq:extrpart}
is pretty straightforward. Indeed 
\begin{eqnarray}
&&\left.\frac{\partial \log\left| \mathcal{Z}_{U(N_c)_1,X} (\Delta,N_c)\right|}{\partial \Delta}
\right|_{\Delta=\frac{1}{N_c+1}} =
\sum_{j=1}^{N_c} \left(\frac{j(N_c+1-j)}{N_c+1}\right) \cot\left(\frac{j \, \pi }{N_c+1}\right)
= \nonumber \\
=&& \frac{1}{2} \left(
\sum_{j=1}^{N_c} \left(\frac{j(N_c+1-j)}{N_c+1}\right) \cot\left(\frac{j \, \pi }{N_c+1}\right)
+
\left(j \rightarrow N_c+1-j'\right) 
\right)=0
\end{eqnarray}
It follows that 
the singlets do not have the canonical scaling dimension, and there are
 $\left[\frac{N_c+1}{2}\right]$ gauge invariant operators with  $R$-charge below or
at the unitarity bound; thus, we have to treat them as free fields, and modify the extremization
principle according to equation \eqref{eq:modifiedF}. We interpret this first step by noticing that
along the RG flow, the operators ${\rm Tr}X^j$ with $j<\left[\frac{N_c+1}{2}\right]$
 will hit the unitarity bound at higher energies.
For high enough $N_c$, this is not the end of the story: extremization of the modified free energy
shows that we did not cure all the accidental symmetries. Again, roughly half of the operators have
$R$-charges below or at the unitarity bound, and we again apply \eqref{eq:modifiedF}.
\footnote{More
precisely, the number of operators is $\left[\frac{N_c+2}{4}\right]$ for $N_c$ even and
$\left[\frac{N_c+1}{4}\right]$ for $N_c$ odd, and the solution is
 $\Delta=\frac{2}{3N_c+2}$
and $\Delta=\frac{2}{3(N_c+1)}$ for $N_c$ even and odd respectively.
These formulas  can be proved by induction. Since a proof would be very marginal to our discussion,
we do not include it in this paper.}
The process
continues until all but one operator, namely $u_{N_c}$, remains and we end up with the following
modified partition function
\bea
\left|\mathcal{Z}_{U(N_c)_1,X}\right| = 2^{-\frac{N_c-1}{2}} e^{l(1- N_c \Delta)}
\eea
which is extremized at $N_c \Delta = 1/2$. We have then shown that the $U(N_c)_1$
partition function coincides with the one of $N_c$ free fields $u_j$, and that a proper treatment
of the accidental symmetries allows us to identify the duality map as $u_j={\rm Tr} X^j$.
The same arguments may be carried over to other models.

\section{Open questions} \label{sec:open}

We provided some nontrivial evidence for classes of infinite three-dimensional dualities for theories with
unitary, orthogonal and symplectic gauge groups. Our results provide support for arbitrary
gauge group and CS levels, and extend previous results which were limited either to
the large-$N$ limit or to numerical evaluations for low ranks and one factor in the gauge group.

Our main tool has been the exact, all-loop partition function evaluated on a squashed three sphere.
Allowing for arbitrary $R$-charges, it can be written as an integral of hyperbolic  functions
which have been recently studied by mathematicians. 

Exact evaluation of the above quantities, available in the literature for classical gauge groups,
allowed to uncover new dualities.
In the large-$N$ limit, and for low enough CS levels, they could also be inferred by
the AdS/CFT duality, and exact evaluation of
the above quantities allows for an extension to arbitrary ranks and levels.

Unitary gauge groups have been extensively studied in the large-$N$ limit, and precise prescriptions
for the computation of the partition function in this regime are available in the literature. Because of its
simplifying nature, it is much more tractable than the computation of the finite-$N$ partition function
and it  allows for comparison of physical quantities in the AdS/CFT correspondence. Based on this
observation, we tried analyzing the case of the other classical gauge groups, where a
similar analysis still lacks. We found that the set of saddle point equations are not consistent with
the long range cancellation in these cases. Thus, the continuum limit would require a different approach.
A similar situation also holds in chiral-like models for unitary gauge groups \cite{Amariti:2011jp,Amariti:2011uw}.
There exist other dualities between 
quivers with unitary gauge groups and quivers
with symplectic/orthogonal gauge groups \cite{Aharony:2008gk}.
These dualities suggest that the theories with symplectic and orthogonal groups
also exhibit the $N^{3/2}$ scaling of the free energy at large $N$. It will be interesting to 
prove the matching of the partition function
for these dualities along the lines of this paper.

Some of the dualities we have studied involve free field theories on the magnetic side. 
Some comments are in order.
Any nontrivial check involving the partition function in this case requires the possibility of an exact
evaluation of the full matrix integral, because on the free theory side there is no integral at all. Secondly,
when one considers such theories, it turns out that the free theory contains $n$ free fields with charge
$j \Delta$, with $\Delta$ the smallest charge and $j=1,\ldots,n$. While this constitutes an offshell check of the duality,
we know that a free field has $R$-charge $1/2$, which cannot be obtained by extremization of the naive
partition function. If the duality holds, this means that the electric $R$-symmetry mixes with an accidental
symmetry and we showed how to handle this scenario in  Section \ref{sec:acc}.

More generally accidental 
symmetries arise  in presence of gauge theories with  tensor matter and superpotential \cite{Kapustin:2011vz}.
These dualities are  three dimensional generalizations  of the  KSS dualities \cite{Kutasov:1995ss}.
It would be interesting to study the matching of the partition functions
in these cases, as already proposed in \cite{Morita:2011cs}, at finite values for the ranks of the gauge groups and CS level.

We conclude by recalling that accidental symmetries are one of  the main issue in the proof of a  $c$-theorem.\footnote{See
\cite{Amariti:2010sz} for other subtleties related to them.}
In the three dimensional case
the candidate $c$-function in
is the free energy $F$ on the round $S^3$ ($F=-\log|\mathcal{Z}|$), which has been 
conjecture to decrease along the RG flow \cite{Jafferis:2011zi}.
Relevant deformations break the abelian symmetries which are manifest in the UV description of the theory
and once we
have a quantity that is maximized by the exact superconformal $R$-symmetry
\footnote{See \cite{Closset:2012vg} for a recent discussion on the maximization of $F$.} we can interpret it as the $c$-function.
The $c$-theorem immediately follows from the two line "almost proof" of \cite{Intriligator:2003jj}.
However accidental symmetries constitute a 
loophole to this argument and a proof of the $F$-theorem requires 
more care in this case: the free field value is a maximum for the function $-\ell(1-\Delta)$,
 thus
the infrared correction term in \eqref{eq:modifiedF}
is always positive, for any value of the scaling dimensions, in full agreement
with the maximization of $F$. However,
the correction term adds a positive contribution to
$F_{IR}$, possibly invalidating the $F$-theorem $F_{IR}<F_{UV}$.

\section*{Acknowledgments}
It is a pleasure to thank
Ofer Aharony,
Francesco Benini,
Cyril Closset,
Kenneth Intriligator,
Claudius Klare,
Alberto Mariotti,
Alessandro Tomasiello and
Alberto Zaffaroni
for interesting discussions and comments.
P.~A.~ and A.~ A.~ are supported by UCSD grant DOE-FG03-97ER40546.
M.~S.~ is a Feinberg Postdoctoral Fellow at the Weizmann Institute of Science.

\appendix

\section{Relations among hyperbolic integrals} \label{vdb section}

In this appendix we review the  equivalence among the hyperbolic integrals 
necessary to match the dual phases in the quiver gauge 
theories that we studied in the paper.
We refer to \cite{fvdb} for more details.

\subsection{The unitary case}

The partition function for a $U(n)$ gauge theory with CS level $2 t$, $s_1$ fundamentals,  $s_2$ anti-fundamentals and one adjoint matter field
 corresponds to the integral  dubbed as $\textit{JI}_{n,(s_1,s_2),t} (\mu;\nu;\lambda;\tau)$ in \cite{fvdb}. 
The original integral is defined as
\begin{align*}
\textit{JI}_{n,(s_1,s_2),t}(\mu;\nu;\lambda;\tau) & =  \frac{\Gamma_h(\tau)^n}{\sqrt{-\omega_1\omega_2}^n n!} \int \prod_{i \leq j < k \leq n}\frac{\Gamma_h(\tau\pm (x_j-x_k))}{\Gamma_h(\pm (x_j-x_k))} \\
 & \times \prod_{j=1}^n \prod_{r=1}^{s_1} \Gamma_h(\mu_r - x_j)\prod_{s=1}^{s_2} \Gamma_h(\nu_s + x_j)c(2 \lambda x_j + t x_j^2) dx_j
\end{align*}
The  variables $\tau$, $\nu$ and $\mu$  are linear combinations of the chemical potentials  
for the global symmetries under which the adjoint, fundamental and anti-fundamental fields are 
 charged respectively.

In the cases studied in section \ref{sec:dualities}
the theory does not contain an adjoint.
This corresponds to identifying the parameter $\tau$ with $\omega$. 
In the hyperbolic  function analysis, setting $\tau=\omega$, removes the  adjoint field contributions from the above integral
because of  (\ref{refle}) and (\ref{otherid}). 
The new integral is defined as
\begin{equation} \label{intJ}
\textit{J}_{n,(s_1,s_2),t} (\mu;\nu;\lambda) = \textit{JI}_{n,(s_1,s_2),t} (\mu;\nu;\lambda;\omega)
\end{equation}
The field theory duality is translated in an equivalence between the  integrals in  (\ref{intJ}).
These equivalences are derived from the 
transformation properties of certain integrals 
named \emph{degenerations} in \cite{fvdb}
\begin{equation}
\label{praritgay}
\textit{I}_{n,\xi}^{m}(\mu;\nu;\lambda)=\textit{J}_{n,(s_1,s_2),t} (\mu;\nu;\lambda)
\end{equation}
where $\xi$ labels the integrals on the LHS of (\ref{praritgay}). The value taken by 
$\xi$ is
either (p,q)$a$ or (p,q)$b$ and it can be fixed by using the following table
\begin{center}
\begin{tabular}{l|c|c|c|c}
condition & type & $m$ & p  &  q  \\
\hline
$t <- |s_1-s_2|$ & $ \text{(p,q)}a $ & $\frac{s_1+s_2-t-2n}{2}$ &  $\frac{s_1- s_2-t+4}{2}$   &  $\frac{s_2- s_1-t+4}{2}$ \\
\hline
$t >  |s_1-s_2|$ & $ \text{(p,q)}b $ & $\frac{s_1+s_2+t-2n}{2}$ &  $\frac{s_2- s_1+t+4}{2}$   &  $\frac{s_1- s_2+t+4}{2}$ 
\end{tabular}
\end{center}
Even if  the definition of $\textit{I}_{n,\xi}^{m}$ looks like a  re-parametrization of  $\textit{J}_{n,(s_1,s_2),t}\,$, the equality 
(\ref{praritgay})
is valid only under certain very \emph{broad}  conditions on the $\mu$, $\nu$ and $\tau$ variables
\footnote{We can always suppose that the values of $\mu$, $\nu$ and $\tau$
are quite generic and that this does not spoil the relations between the integrals.}
. 
At this point of the discussion we prefer to switch to more physical notations, that  involve the usual terminology for the gauge group ranks, the CS level 
and the number of flavors.
Thus  the quantities $n$, $m$, $s_1$, $s_2$ and $t$ are redefined as
\begin{equation}
\label{physvar}
n= N_c
\quad,\quad
m= \widetilde N_c
\quad,\quad
s_1= N_f
\quad,\quad
s_1= \widetilde N_f
\quad,\quad
t =-2 k
\quad,\quad
\end{equation}
We are only interested in non chiral like theories and therefore 
 fix $\widetilde N_f =  N_f$.
In terms of these variables the table becomes
\begin{center}
\begin{tabular}{l|c|c|c|c}
condition & type & $\widetilde N_c$ & p  &  q  \\
\hline
$k > 0$ & $ \text{(p,q)}a $ & $N_c +k $ &  $2+k$   &  $2+k$ \\
\hline
$k < 0$ & $ \text{(p,q})b $ & $N_c -k $ &  $2-k$   &  $2-k$ 
\end{tabular}
\end{center}
Eventually the most useful result of \cite{fvdb}, for our applications, is that the  $a$ and $b$ type integrals are related as
\footnote{As observed in \cite{Benini:2011mf} this result slightly differs from the one on \cite{fvdb}. We are grateful to 
the authors of \cite{Benini:2011mf} for discussions on this point.}
 \begin{eqnarray}
\textit{I}_{n,(p,q)a}^m (\mu;\nu;\lambda) & =&\textit{I}_{m,(p,q)b}^n (\omega-\nu;\omega-\mu;(p-q)\omega-\lambda) \prod_{r,s}\Gamma_h(\mu_r+\nu_s)\zeta^{(-6+2p+2q-pq)}
\nonumber
\\
& \times& c((\frac{1}{2}(p-q)^2+pq+(4-p-q)(m+2)-4)\omega^2+\frac{1}{2}\lambda^2)\\
&\times& c((2-p)\sum_r\mu_r^2+(2-q)\sum_s\nu_s^2+\frac{1}{2}(2m\omega -\sum_r\mu_r-\sum_s\nu_s)^2)\nonumber\\
& \times& c(\lambda(\sum_r\mu_r-\sum_s\nu_s+(p-q)\omega)+(p+q-4)(\sum_r\mu_r+\sum_s\nu_s)\omega)\nonumber
\end{eqnarray}
where $r=1,\dots,m+n+2-q\,(\equiv s_1)$ and $s=1,\dots,m+n-2+p\,(\equiv s_2)$.
Upon substituting (\ref{physvar})  and   fixing $\widetilde N_f=N_f$ this becomes
\small
\begin{eqnarray}
\label{Ucase}
\textit{I}_{N_c,(2+k,2+k)a}^{\widetilde N_c} (\mu;\nu;\lambda)\!&& \!=\textit{I}_{\tilde N_c,(2+k,2+k)b}^{N_c} (\omega-\nu;\omega-\mu;-\lambda) 
\prod_{r,s=1}^{N_f} \Gamma_h(\mu_r+\nu_s)\zeta^{-k^2-2}\nonumber\\
&& \times c\left(
k\left(\!\sum_{r=1}^{N_f} \mu_r^2 \!+\!\sum_{s=1}^{N_f} \nu_r^2\!\right)\!+\! 
k(k-2m)\omega^2+\frac{1}{2}\lambda^2
\!-\!2 k\left(\!\sum_{r=1}^{N_f} \mu_r\!+\!\sum_{s=1}^{N_f} \nu_s\!\right)\omega\!\right)
 \nonumber\\
&& \times c\left(\lambda\left(\sum_{r=1}^{N_f} \mu_r-\sum_{s=1}^{N_f}\nu_s\right)+
\frac{1}{2}(2m\omega -\sum_{r=1}^{N_f}\mu_r-\sum_{s=1}^{N_f} \nu_s)^2\right)
\end{eqnarray}
\normalsize
 A few comments are in order. First  the difference between the case $a$ and $b$  is in the sign of the CS level $k$.
 In this case we fixed $k>0$ but the same equality can be reversed if one starts with $k<0$ and use the equation
(5.5.7) in \cite{fvdb}. This identifies $\tilde N_c$ with $N_c+|k|$.  
Moreover, as discussed in \cite{fvdb},  $t+s_1+s_2$ is always even for the above  \emph{degenerations}.  
 This  corresponds to requiring   $|k|+\frac{N_f+\tilde N_f}{2}$ to be integer.
 This is the same as the  parity anomaliy condition of three dimensional field theories  \cite{Niemi:1983rq}.

\subsection{The symplectic case}

The second class of integral that we need from \cite{fvdb}
is associated with the symplectic group $SP(2N_c)_{k}$.
The integrals have been dubbed  as 
 $\textit{JI}_{n,s,t}(\mu; \tau)$ in \cite{fvdb}.  Explicitly they are  
 \small
\begin{eqnarray}
\label{ref:SPvdb}
\textit{JI}_{n,s_1,t}(\mu; \tau)  = \displaystyle
\frac{\Gamma_h(\tau)^n}{\sqrt{-\omega_1\omega_2}^n n!} \int  \frac{\displaystyle 
\prod_{i \leq j < k \leq n}\Gamma_h(\tau\pm x_j\pm x_k) \prod_{j=1}^n \prod_{r=1}^{s_1} \Gamma_h(\mu_r \pm x_j)}{\displaystyle\prod_{i \leq j < k \leq n}\Gamma_h(\pm x_j \pm x_k)  \prod_{j=1}^n \Gamma_h(\pm 2 x_j)} \prod_{j=1}^n c(2 t x_j^2) dx_j \nonumber 
\\
\end{eqnarray} 
\normalsize
In this case  $\tau$  labels the fields in the antisymmetric representation while $\mu$ is the label for fields in the fundamental representation. 
In the absence any anti-symmetric representations $\tau$ gets identified with $\omega$. 
In this case the integral (\ref{ref:SPvdb}) becomes
 \begin{eqnarray} \label{twointvdb}       
 \textit{I}_{n,\text{p}a}^m(\mu)=\textit{JI}_{n,2n+2m+4-\text{p},2-\text{p}}(\mu;\omega) \\
 \textit{I}_{n,\text{p}b}^m(\mu)=\textit{JI}_{n,2n+2m+4-\text{p},\text{p}-2}(\mu;\omega) \nonumber   
 \end{eqnarray}
where 
where p$a$ or p$b$ are fixed as
\begin{center}
\begin{tabular}{l|c|c|c}
condition & type & $m$ & p    \\
\hline
$t <0$ & $ \text{p}a $ & $\frac{s_1-t-2n-2}{2}$ &  $2-t$    \\
\hline
$t >  0$ & $ \text{p}b $ &$\frac{s_1-t-2n-2}{2}$ &  $2+t$  
\end{tabular}
\end{center}
As in the case of unitary groups we switch to more physical parameters
\begin{equation}
t=-2 k\quad n=N_c\quad s_1=2N_f \quad m=\widetilde N_c 
\end{equation}
In temrs of these parameters the table becomes
\begin{center}
\begin{tabular}{l|c|c|c}
condition & type & $\widetilde N_c$ & p    \\
\hline
$k>0$ & $ \text{p}a $ & $N_f+k-N_c-1$ &  $2(1+k)$    \\
\hline
$k<  0$ & $ \text{p}b $ & $N_f-k-N_c-1$&  $2(1-k)$  
\end{tabular}
\end{center}
The transformation properties of these integrals, given in \cite{fvdb},
become (we fix $k>0$)
\small
\begin{eqnarray}\label{SPcase}
\textit{I}_{N_c,2(1+k)a}^{\widetilde N_c}(\mu) & = & \textit{I}_{\widetilde{N_c},2(1+k)b}^{N_c}(\omega-\mu) \prod_{1\leq r<s\leq 2 N_f}\Gamma_h(\mu_r+\mu_s) \zeta^{(k-1)(1-2k)}  \\
&\times
&
 c\left(-2 k \sum_{r=1}^{2 N_f}(\mu_r-\omega)^2+\left(\left(2 \widetilde N_c+1\right)\omega-\sum_{r=1}^{2N_f} \mu_r\right)^2+2 k \left(2 N_c-\frac{2k-1}{2}\right)\omega^2\right)\nonumber 
\end{eqnarray}
\normalsize
 As in the unitary case the difference between the case $a$ and $b$  is in the sign of the CS level $k$,
 and the case with $k<0$ is obtained from (\ref{SPcase}) after using  relation (5.5.2) of \cite{fvdb} .

\section{Characters}
\label{app:A}

In the paper we studied different representations
for the orthogonal, symplectic and unitary groups. In this appendix we 
list the formula for the characters of the representation of these groups.
As usual we identify a representation of a  simple group of rank $n$ by its Dynkin labels,
a set of $n$ integers $(s_1,\dots,s_n)$ which are  assigned to the simple roots of the group
by the Dynkin diagrams.
Then the characters of the representations are associated to the Schur polynomials 
as functions of the eigenvalues of the group $G$, parameterizing the maximal abelian torus.
In the cases we investigated the Schur polynomials are
\begin{itemize}
\item \underline{$U(n)$}
\begin{equation}
P_{\vec s} = \frac{\det{z_i^{s_j+n-j}}}{\det {z_i^{n-j}}} \quad \quad  i,j=1,\dots,n
\end{equation}
\item \underline{$SP(2n)$}
\begin{equation}
P_{\vec s}  = \frac{\det{\left(z_i^{s_j+n-j+1}-z_i^{-(s_j+n-j+1)}\right)}}
{\det \left(z_i^{n-j+1}-z_i^{-(n-j+1)}\right)}
\quad \quad i,j=1,\dots,n
\end{equation}
\item \underline{$SO(2n)$}
\begin{equation}
P_{\vec s}  = \frac{\det{\left(z_i^{s_j+n-j}+z_i^{-(s_j+n-j)}\right)}
+\det{\left(z_i^{s_j+n-j}-z_i^{-(s_j+n-j)}\right)}
}
{2 \det \left(z_i^{n-j+1}-z_i^{-(n-j+1)}\right)}
\prod_{i=1}^{n} \left( z_i-\frac{1}{z_i}\right)
\end{equation}
with $ i,j=1,\dots,n$
\item \underline{$SO(2n+1)$}
\begin{equation}
P_{\vec s}  = 
\frac{\det{\left(z_i^{s_j+\frac{1}{2}+n-j}+z_i^{-(s_j+\frac{1}{2}+n-j)}\right)}}
{2 \det \left(z_i^{n-j+1}-z_i^{-(n-j+1)}\right)}
\frac{\displaystyle  \prod_{i=1}^{n} \left( z_i-\frac{1}{z_i}\right)}
{\displaystyle \prod_{i=1}^{n} \left( z_i^{\frac{1}{2}}-\frac{1}{z_i^{\frac{1}{2}}}\right)}
\end{equation}
\end{itemize}
In the computation of the partition function we actually used the substitution
\begin{equation}
z_i = e^{i x_i}
\end{equation}
and we studied the characters to respect to the $x_i$ variables.
For example in the adjoint representation we
have
\begin{center}
\begin{tabular}{c|c|c}
Group&Dynkin Label& Non Zero Roots $(i<j)$ \\
\hline
$U(n)$ & $s=(2,1,\dots,1,0)$ & $\pm(x_i-x_j)$ \\
$SP(2n)$ & $s=(2,0,\dots,0,0)$ & $\pm x_i \pm x_j$, $\quad \pm 2 x_i$\\
$SO(2n)$ &$s=(1,1,0,\dots,0,0)$ & $\pm x_i \pm x_j$\\
 $SO(2n+1)$ &$s=(1,1,0,\dots,0,0)$ & $\pm x_i \pm x_j$, $\quad \pm x_i$\\
\end{tabular}
\end{center}
In addition in every case there are $n$ zero roots associated to the adjoint of the four cases.
By applying the same formulas we can obtain the characters for the other representations.

\bibliographystyle{JHEP}
\bibliography{duality}

\providecommand{\href}[2]{#2}\begingroup\raggedright\begin{thebibliography}{10}

\bibitem{Aharony:1997gp}
O.~Aharony, {\it {IR duality in d = 3 N=2 supersymmetric USp(2N(c)) and U(N(c))
  gauge theories}},  {\em Phys.Lett.} {\bf B404} (1997) 71--76,
  [\href{http://xxx.lanl.gov/abs/hep-th/9703215}{{\tt hep-th/9703215}}].

\bibitem{Giveon:2008zn}
A.~Giveon and D.~Kutasov, {\it {Seiberg Duality in Chern-Simons Theory}},  {\em
  Nucl.Phys.} {\bf B812} (2009) 1--11,
  [\href{http://xxx.lanl.gov/abs/0808.0360}{{\tt arXiv:0808.0360}}].

\bibitem{Kapustin:2009kz}
A.~Kapustin, B.~Willett, and I.~Yaakov, {\it {Exact Results for Wilson Loops in
  Superconformal Chern-Simons Theories with Matter}},  {\em JHEP} {\bf 1003}
  (2010) 089, [\href{http://xxx.lanl.gov/abs/0909.4559}{{\tt
  arXiv:0909.4559}}].

\bibitem{Jafferis:2010un}
D.~L. Jafferis, {\it {The Exact Superconformal R-Symmetry Extremizes Z}},
  \href{http://xxx.lanl.gov/abs/1012.3210}{{\tt arXiv:1012.3210}}.

\bibitem{Hama:2010av}
N.~Hama, K.~Hosomichi, and S.~Lee, {\it {Notes on SUSY Gauge Theories on
  Three-Sphere}},  {\em JHEP} {\bf 1103} (2011) 127,
  [\href{http://xxx.lanl.gov/abs/1012.3512}{{\tt arXiv:1012.3512}}].

\bibitem{Suyama:2009pd}
T.~Suyama, {\it {On Large N Solution of ABJM Theory}},  {\em Nucl.Phys.} {\bf
  B834} (2010) 50--76, [\href{http://xxx.lanl.gov/abs/0912.1084}{{\tt
  arXiv:0912.1084}}].

\bibitem{Herzog:2010hf}
C.~P. Herzog, I.~R. Klebanov, S.~S. Pufu, and T.~Tesileanu, {\it {Multi-Matrix
  Models and Tri-Sasaki Einstein Spaces}},  {\em Phys.Rev.} {\bf D83} (2011)
  046001, [\href{http://xxx.lanl.gov/abs/1011.5487}{{\tt arXiv:1011.5487}}].

\bibitem{Martelli:2011qj}
D.~Martelli and J.~Sparks, {\it {The large N limit of quiver matrix models and
  Sasaki-Einstein manifolds}},  {\em Phys.Rev.} {\bf D84} (2011) 046008,
  [\href{http://xxx.lanl.gov/abs/1102.5289}{{\tt arXiv:1102.5289}}].

\bibitem{Cheon:2011vi}
S.~Cheon, H.~Kim, and N.~Kim, {\it {Calculating the partition function of N=2
  Gauge theories on $S^3$ and AdS/CFT correspondence}},  {\em JHEP} {\bf 1105}
  (2011) 134, [\href{http://xxx.lanl.gov/abs/1102.5565}{{\tt
  arXiv:1102.5565}}].

\bibitem{Jafferis:2011zi}
D.~L. Jafferis, I.~R. Klebanov, S.~S. Pufu, and B.~R. Safdi, {\it {Towards the
  F-Theorem: N=2 Field Theories on the Three-Sphere}},  {\em JHEP} {\bf 1106}
  (2011) 102, [\href{http://xxx.lanl.gov/abs/1103.1181}{{\tt
  arXiv:1103.1181}}].

\bibitem{Amariti:2011uw}
A.~Amariti, C.~Klare, and M.~Siani, {\it {The Large N Limit of Toric
  Chern-Simons Matter Theories and Their Duals}},
  \href{http://xxx.lanl.gov/abs/1111.1723}{{\tt arXiv:1111.1723}}.

\bibitem{Minwalla:2011ma}
S.~Minwalla, P.~Narayan, T.~Sharma, V.~Umesh, and X.~Yin, {\it {Supersymmetric
  States in Large N Chern-Simons-Matter Theories}},  {\em JHEP} {\bf 1202}
  (2012) 022, [\href{http://xxx.lanl.gov/abs/1104.0680}{{\tt
  arXiv:1104.0680}}].

\bibitem{Amariti:2011da}
A.~Amariti and M.~Siani, {\it {Z-extremization and F-theorem in Chern-Simons
  matter theories}},  {\em JHEP} {\bf 1110} (2011) 016,
  [\href{http://xxx.lanl.gov/abs/1105.0933}{{\tt arXiv:1105.0933}}].

\bibitem{Amariti:2011jp}
A.~Amariti and M.~Siani, {\it {Z Extremization in Chiral-Like Chern Simons
  Theories}},  \href{http://xxx.lanl.gov/abs/1109.4152}{{\tt arXiv:1109.4152}}.

\bibitem{Gang:2011jj}
D.~Gang, C.~Hwang, S.~Kim, and J.~Park, {\it {Tests of AdS$_4$/CFT$_3$
  correspondence for $\mathcal{N}=2$ chiral-like theory}},  {\em JHEP} {\bf
  1202} (2012) 079, [\href{http://xxx.lanl.gov/abs/1111.4529}{{\tt
  arXiv:1111.4529}}].

\bibitem{Amariti:2011xp}
A.~Amariti and M.~Siani, {\it {F-maximization along the RG flows: A Proposal}},
   {\em JHEP} {\bf 1111} (2011) 056,
  [\href{http://xxx.lanl.gov/abs/1105.3979}{{\tt arXiv:1105.3979}}].

\bibitem{Drukker:2010nc}
N.~Drukker, M.~Marino, and P.~Putrov, {\it {From weak to strong coupling in
  ABJM theory}},  {\em Commun.Math.Phys.} {\bf 306} (2011) 511--563,
  [\href{http://xxx.lanl.gov/abs/1007.3837}{{\tt arXiv:1007.3837}}].

\bibitem{Amariti:2012tj}
A.~Amariti and S.~Franco, {\it {Free Energy vs Sasaki-Einstein Volume for
  Infinite Families of M2-Brane Theories}},
  \href{http://xxx.lanl.gov/abs/1204.6040}{{\tt arXiv:1204.6040}}.

\bibitem{Bagger:2006sk}
J.~Bagger and N.~Lambert, {\it {Modeling Multiple M2's}},  {\em Phys.Rev.} {\bf
  D75} (2007) 045020, [\href{http://xxx.lanl.gov/abs/hep-th/0611108}{{\tt
  hep-th/0611108}}]. Dedicated to the Memory of Andrew Chamblin.

\bibitem{Bagger:2007jr}
J.~Bagger and N.~Lambert, {\it {Gauge symmetry and supersymmetry of multiple
  M2-branes}},  {\em Phys.Rev.} {\bf D77} (2008) 065008,
  [\href{http://xxx.lanl.gov/abs/0711.0955}{{\tt arXiv:0711.0955}}].

\bibitem{Gustavsson:2007vu}
A.~Gustavsson, {\it {Algebraic structures on parallel M2-branes}},  {\em
  Nucl.Phys.} {\bf B811} (2009) 66--76,
  [\href{http://xxx.lanl.gov/abs/0709.1260}{{\tt arXiv:0709.1260}}].

\bibitem{Bagger:2007vi}
J.~Bagger and N.~Lambert, {\it {Comments on multiple M2-branes}},  {\em JHEP}
  {\bf 0802} (2008) 105, [\href{http://xxx.lanl.gov/abs/0712.3738}{{\tt
  arXiv:0712.3738}}].

\bibitem{Aharony:2008ug}
O.~Aharony, O.~Bergman, D.~L. Jafferis, and J.~Maldacena, {\it {N=6
  superconformal Chern-Simons-matter theories, M2-branes and their gravity
  duals}},  {\em JHEP} {\bf 0810} (2008) 091,
  [\href{http://xxx.lanl.gov/abs/0806.1218}{{\tt arXiv:0806.1218}}].

\bibitem{Gaiotto:2007qi}
D.~Gaiotto and X.~Yin, {\it {Notes on superconformal Chern-Simons-Matter
  theories}},  {\em JHEP} {\bf 0708} (2007) 056,
  [\href{http://xxx.lanl.gov/abs/0704.3740}{{\tt arXiv:0704.3740}}].

\bibitem{Avdeev:1992jt}
L.~Avdeev, D.~Kazakov, and I.~Kondrashuk, {\it {Renormalizations in
  supersymmetric and nonsupersymmetric nonAbelian Chern-Simons field theories
  with matter}},  {\em Nucl.Phys.} {\bf B391} (1993) 333--357.

\bibitem{Jafferis:2008qz}
D.~L. Jafferis and A.~Tomasiello, {\it {A Simple class of N=3 gauge/gravity
  duals}},  {\em JHEP} {\bf 0810} (2008) 101,
  [\href{http://xxx.lanl.gov/abs/0808.0864}{{\tt arXiv:0808.0864}}].

\bibitem{Martelli:2008si}
D.~Martelli and J.~Sparks, {\it {Moduli spaces of Chern-Simons quiver gauge
  theories and AdS(4)/CFT(3)}},  {\em Phys.Rev.} {\bf D78} (2008) 126005,
  [\href{http://xxx.lanl.gov/abs/0808.0912}{{\tt arXiv:0808.0912}}].

\bibitem{Hanany:2008cd}
A.~Hanany and A.~Zaffaroni, {\it {Tilings, Chern-Simons Theories and M2
  Branes}},  {\em JHEP} {\bf 0810} (2008) 111,
  [\href{http://xxx.lanl.gov/abs/0808.1244}{{\tt arXiv:0808.1244}}].

\bibitem{Hanany:2008fj}
A.~Hanany, D.~Vegh, and A.~Zaffaroni, {\it {Brane Tilings and M2 Branes}},
  {\em JHEP} {\bf 0903} (2009) 012,
  [\href{http://xxx.lanl.gov/abs/0809.1440}{{\tt arXiv:0809.1440}}].

\bibitem{Ueda:2008hx}
K.~Ueda and M.~Yamazaki, {\it {Toric Calabi-Yau four-folds dual to
  Chern-Simons-matter theories}},  {\em JHEP} {\bf 0812} (2008) 045,
  [\href{http://xxx.lanl.gov/abs/0808.3768}{{\tt arXiv:0808.3768}}].

\bibitem{Imamura:2008nn}
Y.~Imamura and K.~Kimura, {\it {On the moduli space of elliptic
  Maxwell-Chern-Simons theories}},  {\em Prog.Theor.Phys.} {\bf 120} (2008)
  509--523, [\href{http://xxx.lanl.gov/abs/0806.3727}{{\tt arXiv:0806.3727}}].

\bibitem{Franco:2008um}
S.~Franco, A.~Hanany, J.~Park, and D.~Rodriguez-Gomez, {\it {Towards M2-brane
  Theories for Generic Toric Singularities}},  {\em JHEP} {\bf 0812} (2008)
  110, [\href{http://xxx.lanl.gov/abs/0809.3237}{{\tt arXiv:0809.3237}}].

\bibitem{Hanany:2008gx}
A.~Hanany and Y.-H. He, {\it {M2-Branes and Quiver Chern-Simons: A Taxonomic
  Study}},  \href{http://xxx.lanl.gov/abs/0811.4044}{{\tt arXiv:0811.4044}}.

\bibitem{Bianchi:2009ja}
M.~S. Bianchi, S.~Penati, and M.~Siani, {\it {Infrared stability of ABJ-like
  theories}},  {\em JHEP} {\bf 1001} (2010) 080,
  [\href{http://xxx.lanl.gov/abs/0910.5200}{{\tt arXiv:0910.5200}}].

\bibitem{Amariti:2009rb}
A.~Amariti, D.~Forcella, L.~Girardello, and A.~Mariotti, {\it {3D Seiberg-like
  Dualities and M2 Branes}},  {\em JHEP} {\bf 1005} (2010) 025,
  [\href{http://xxx.lanl.gov/abs/0903.3222}{{\tt arXiv:0903.3222}}].

\bibitem{Franco:2009sp}
S.~Franco, I.~R. Klebanov, and D.~Rodriguez-Gomez, {\it {M2-branes on Orbifolds
  of the Cone over Q**1,1,1}},  {\em JHEP} {\bf 0908} (2009) 033,
  [\href{http://xxx.lanl.gov/abs/0903.3231}{{\tt arXiv:0903.3231}}].

\bibitem{Bianchi:2009rf}
M.~S. Bianchi, S.~Penati, and M.~Siani, {\it {Infrared Stability of N = 2
  Chern-Simons Matter Theories}},  {\em JHEP} {\bf 1005} (2010) 106,
  [\href{http://xxx.lanl.gov/abs/0912.4282}{{\tt arXiv:0912.4282}}].

\bibitem{Davey:2009sr}
J.~Davey, A.~Hanany, N.~Mekareeya, and G.~Torri, {\it {Phases of M2-brane
  Theories}},  {\em JHEP} {\bf 0906} (2009) 025,
  [\href{http://xxx.lanl.gov/abs/0903.3234}{{\tt arXiv:0903.3234}}].

\bibitem{Kapustin:2010mh}
A.~Kapustin, B.~Willett, and I.~Yaakov, {\it {Tests of Seiberg-like Duality in
  Three Dimensions}},  \href{http://xxx.lanl.gov/abs/1012.4021}{{\tt
  arXiv:1012.4021}}.

\bibitem{Kapustin:2010xq}
A.~Kapustin, B.~Willett, and I.~Yaakov, {\it {Nonperturbative Tests of
  Three-Dimensional Dualities}},  {\em JHEP} {\bf 1010} (2010) 013,
  [\href{http://xxx.lanl.gov/abs/1003.5694}{{\tt arXiv:1003.5694}}].

\bibitem{Gulotta:2011vp}
D.~R. Gulotta, J.~Ang, and C.~P. Herzog, {\it {Matrix Models for Supersymmetric
  Chern-Simons Theories with an ADE Classification}},  {\em JHEP} {\bf 1201}
  (2012) 132, [\href{http://xxx.lanl.gov/abs/1111.1744}{{\tt
  arXiv:1111.1744}}].

\bibitem{Hama:2011ea}
N.~Hama, K.~Hosomichi, and S.~Lee, {\it {SUSY Gauge Theories on Squashed
  Three-Spheres}},  {\em JHEP} {\bf 1105} (2011) 014,
  [\href{http://xxx.lanl.gov/abs/1102.4716}{{\tt arXiv:1102.4716}}].

\bibitem{fvdb}
F.~van~de Bult, {\it {Hyperbolic Hypergeometric Functions,
  http://www.its.caltech.edu/~vdbult/Thesis.pdf}},  {\em Thesis} (2008).

\bibitem{Willett:2011gp}
B.~Willett and I.~Yaakov, {\it {N=2 Dualities and Z Extremization in Three
  Dimensions}},  \href{http://xxx.lanl.gov/abs/1104.0487}{{\tt
  arXiv:1104.0487}}.

\bibitem{Kapustin:2011gh}
A.~Kapustin, {\it {Seiberg-like duality in three dimensions for orthogonal
  gauge groups}},  \href{http://xxx.lanl.gov/abs/1104.0466}{{\tt
  arXiv:1104.0466}}.

\bibitem{Benini:2011mf}
F.~Benini, C.~Closset, and S.~Cremonesi, {\it {Comments on 3d Seiberg-like
  dualities}},  {\em JHEP} {\bf 1110} (2011) 075,
  [\href{http://xxx.lanl.gov/abs/1108.5373}{{\tt arXiv:1108.5373}}].

\bibitem{Niarchos:2012ah}
V.~Niarchos, {\it {Seiberg dualities and the 3d/4d connection}},
  \href{http://xxx.lanl.gov/abs/1205.2086}{{\tt arXiv:1205.2086}}.

\bibitem{Jafferis:2011ns}
D.~Jafferis and X.~Yin, {\it {A Duality Appetizer}},
  \href{http://xxx.lanl.gov/abs/1103.5700}{{\tt arXiv:1103.5700}}.

\bibitem{Kapustin:2011vz}
A.~Kapustin, H.~Kim, and J.~Park, {\it {Dualities for 3d Theories with Tensor
  Matter}},  {\em JHEP} {\bf 1112} (2011) 087,
  [\href{http://xxx.lanl.gov/abs/1110.2547}{{\tt arXiv:1110.2547}}].

\bibitem{Ruijsenaars}
S.~Ruijsenaars, {\it {First order analytic difference equations and integrable
  quantum systems}},  {\em J. Math. Phys.} {\bf 38} (1997) 1069�1146.

\bibitem{Hanany:1996ie}
A.~Hanany and E.~Witten, {\it {Type IIB superstrings, BPS monopoles, and
  three-dimensional gauge dynamics}},  {\em Nucl.Phys.} {\bf B492} (1997)
  152--190, [\href{http://xxx.lanl.gov/abs/hep-th/9611230}{{\tt
  hep-th/9611230}}].

\bibitem{Hosomichi:2008jb}
K.~Hosomichi, K.-M. Lee, S.~Lee, S.~Lee, and J.~Park, {\it {N=5,6
  Superconformal Chern-Simons Theories and M2-branes on Orbifolds}},  {\em
  JHEP} {\bf 0809} (2008) 002, [\href{http://xxx.lanl.gov/abs/0806.4977}{{\tt
  arXiv:0806.4977}}].

\bibitem{Aharony:2008gk}
O.~Aharony, O.~Bergman, and D.~L. Jafferis, {\it {Fractional M2-branes}},  {\em
  JHEP} {\bf 0811} (2008) 043, [\href{http://xxx.lanl.gov/abs/0807.4924}{{\tt
  arXiv:0807.4924}}].

\bibitem{Armoni:2008kr}
A.~Armoni and A.~Naqvi, {\it {A Non-Supersymmetric Large-N 3D CFT And Its
  Gravity Dual}},  {\em JHEP} {\bf 0809} (2008) 119,
  [\href{http://xxx.lanl.gov/abs/0806.4068}{{\tt arXiv:0806.4068}}].

\bibitem{Forcella:2009jj}
D.~Forcella and A.~Zaffaroni, {\it {N=1 Chern-Simons theories, orientifolds and
  Spin(7) cones}},  {\em JHEP} {\bf 1005} (2010) 045,
  [\href{http://xxx.lanl.gov/abs/0911.2595}{{\tt arXiv:0911.2595}}].

\bibitem{Bhattacharya:2008bja}
J.~Bhattacharya and S.~Minwalla, {\it {Superconformal Indices for N = 6 Chern
  Simons Theories}},  {\em JHEP} {\bf 0901} (2009) 014,
  [\href{http://xxx.lanl.gov/abs/0806.3251}{{\tt arXiv:0806.3251}}].

\bibitem{Kim:2009wb}
S.~Kim, {\it {The Complete superconformal index for N=6 Chern-Simons theory}},
  {\em Nucl.Phys.} {\bf B821} (2009) 241--284,
  [\href{http://xxx.lanl.gov/abs/0903.4172}{{\tt arXiv:0903.4172}}].

\bibitem{Imamura:2011su}
Y.~Imamura and S.~Yokoyama, {\it {Index for three dimensional superconformal
  field theories with general R-charge assignments}},  {\em JHEP} {\bf 1104}
  (2011) 007, [\href{http://xxx.lanl.gov/abs/1101.0557}{{\tt
  arXiv:1101.0557}}].

\bibitem{Goddard:1976qe}
P.~Goddard, J.~Nuyts, and D.~I. Olive, {\it {Gauge Theories and Magnetic
  Charge}},  {\em Nucl.Phys.} {\bf B125} (1977) 1.

\bibitem{Kutasov:2003iy}
D.~Kutasov, A.~Parnachev, and D.~A. Sahakyan, {\it {Central charges and U(1)(R)
  symmetries in N=1 superYang-Mills}},  {\em JHEP} {\bf 0311} (2003) 013,
  [\href{http://xxx.lanl.gov/abs/hep-th/0308071}{{\tt hep-th/0308071}}].

\bibitem{Barnes:2004jj}
E.~Barnes, K.~A. Intriligator, B.~Wecht, and J.~Wright, {\it {Evidence for the
  strongest version of the 4d a-theorem, via a-maximization along RG flows}},
  {\em Nucl.Phys.} {\bf B702} (2004) 131--162,
  [\href{http://xxx.lanl.gov/abs/hep-th/0408156}{{\tt hep-th/0408156}}].

\bibitem{Kutasov:2003ux}
D.~Kutasov, {\it {New results on the 'a theorem' in four-dimensional
  supersymmetric field theory}},
  \href{http://xxx.lanl.gov/abs/hep-th/0312098}{{\tt hep-th/0312098}}.

\bibitem{Morita:2011cs}
T.~Morita and V.~Niarchos, {\it {F-theorem, duality and SUSY breaking in
  one-adjoint Chern-Simons-Matter theories}},  {\em Nucl.Phys.} {\bf B858}
  (2012) 84--116, [\href{http://xxx.lanl.gov/abs/1108.4963}{{\tt
  arXiv:1108.4963}}].

\bibitem{Kutasov:1995ss}
D.~Kutasov, A.~Schwimmer, and N.~Seiberg, {\it {Chiral rings, singularity
  theory and electric - magnetic duality}},  {\em Nucl.Phys.} {\bf B459} (1996)
  455--496, [\href{http://xxx.lanl.gov/abs/hep-th/9510222}{{\tt
  hep-th/9510222}}].

\bibitem{Amariti:2010sz}
A.~Amariti, L.~Girardello, A.~Mariotti, and M.~Siani, {\it {Metastable Vacua in
  Superconformal SQCD-like Theories}},  {\em JHEP} {\bf 1102} (2011) 092,
  [\href{http://xxx.lanl.gov/abs/1003.0523}{{\tt arXiv:1003.0523}}].

\bibitem{Closset:2012vg}
C.~Closset, T.~T. Dumitrescu, G.~Festuccia, Z.~Komargodski, and N.~Seiberg,
  {\it {Contact Terms, Unitarity, and F-Maximization in Three-Dimensional
  Superconformal Theories}},  \href{http://xxx.lanl.gov/abs/1205.4142}{{\tt
  arXiv:1205.4142}}.

\bibitem{Intriligator:2003jj}
K.~A. Intriligator and B.~Wecht, {\it {The Exact superconformal R symmetry
  maximizes a}},  {\em Nucl.Phys.} {\bf B667} (2003) 183--200,
  [\href{http://xxx.lanl.gov/abs/hep-th/0304128}{{\tt hep-th/0304128}}].

\bibitem{Niemi:1983rq}
A.~Niemi and G.~Semenoff, {\it {Axial Anomaly Induced Fermion Fractionization
  and Effective Gauge Theory Actions in Odd Dimensional Space-Times}},  {\em
  Phys.Rev.Lett.} {\bf 51} (1983) 2077.

\end{thebibliography}\endgroup

\end{document}